\documentclass{aa}
\usepackage{graphicx}
\usepackage{epsfig}
\usepackage{txfonts}
\usepackage{natbib}
\usepackage{aalongtable}
\begin{document}
   \title{X-ray absorption in distant type II 
          QSOs\thanks{Based on observations obtained with {\it XMM-Newton}, 
          an ESA science mission with instruments and contributions 
          directly funded by ESA Member States and NASA.}}

   \author{M. Krumpe
          \inst{1}
          \and
          G. Lamer
          \inst{1}
          \and
          A. Corral
          \inst{2}
          \and
          A.D. Schwope
          \inst{1}
          \and
          F.J. Carrera
          \inst{2}
          \and
          X. Barcons
          \inst{2}
          \and
          M. Page
          \inst{3}
          \and
          S. Mateos
          \inst{4}
          \and
          J.A. Tedds
          \inst{4}
          \and
          M.G. Watson
          \inst{4}
          }

   \offprints{M. Krumpe}

   \institute{Astrophysikalisches Institut Potsdam,
              An der Sternwarte 16,
              14482 Potsdam, Germany\\
              \email{mkrumpe@aip.de}
              \and
              Instituto de F\'isica de Cantabria (CSIC-UC), 
              Avenida de los Castros, 39005 Santander, Spain
              \and 
              Mullard Space Science Laboratory,
              Holmbury St. Mary,
              Dorking, United Kingdom
              \and
              Department of Physics \& Astronomy,
              University of Leicester,
              Leicester LE1 7RH, United Kingdom
              }
   \date{Received ; accepted}


  \abstract{}{We present the results of the X-ray spectral analysis of an 
              {\it XMM-Newton}-selected type II
              QSO sample with $z \ge 0.5$ and 0.5-10\,keV flux of
              $0.3-33\times10^{-14}$\,erg/s/cm$^2$. 
              The distribution of absorbing column densities
              in type II QSOs is investigated and the dependence of absorption
              on X-ray luminosity and redshift is studied.}
             {We inspected 51 spectroscopically classified type II QSO candidates from 
              the {\it XMM-Newton} \rm Marano field survey, the 
              {\it XMM-Newton}-2dF wide angle survey (XWAS), and the AXIS 
              survey to set-up a well-defined sample with secure optical
              type II identifications. Fourteen type II QSOs were classified
              and an X-ray spectral analysis performed. Since most of our 
              sources have only $\sim$40 X-ray counts (PN-detector), we carefully studied 
              the fit results of the simulated X-ray spectra as a function of fit statistic
              and binning method. We determined that fitting the spectra with the
              Cash-statistic and a binning of minimum one count per bin
              recovers the input values of the simulated X-ray spectra best. 
              Above 100 PN counts, the free fits of the spectrum's slope and absorbing
              hydrogen column density are reliable. }
             {We find only moderate absorption ($N_{\rm  H}=(2-10)\times10^{22}$\,cm$^{-2}$) 
              and no obvious trends with redshift and intrinsic
              X-ray luminosity. In a few cases a Compton-thick absorber cannot
              be excluded. Two type II objects with no X-ray absorption
              were discovered. We find no evidence for an intrinsic separation between
              type II AGN and high X-ray luminosity type II QSO in terms
              of absorption.
              The stacked X-ray spectrum of our 14 type II QSOs shows no
              iron K$\alpha$ line. In contrast, the stack of
              the 8 type II AGN reveals a very prominent iron K$\alpha$ line
              at an energy of $\sim 6.6$\,keV and an $EW \sim 2$\,keV.}{}


   \maketitle
%

\section{Introduction\label{intro}}

The unification model for active galaxy nuclei (AGN) 
is based on the assumption that the Seyfert type I/type II dichotomy 
in AGN is a result of varying orientation relative to the line of sight 
of similar objects (\citealt{antonucci}). All AGN consist of a
super-massive black hole with an accretion disk. This central engine 
is surrounded by optically thick, toroidally concentrated 
dusty material. An observer can either view the central engine 
directly (type I AGN) or through the optically thick torus 
(type II AGN). Direct observation of the broad line region reveals a
strong UV continuum and broad permitted emission lines while 
the optical spectra of type II AGN show narrow permitted and forbidden 
emission lines. \cite{hao} provided a formal separation criterion 
by finding a bimodal distribution in the H$\alpha$-FWHM with a significant 
dip at FWHM(H$\alpha$)=1200\,km/s.

\cite{kleinmann} discovered the first type II quasi stellar object (QSO).
QSOs show higher intrinsic luminosity than AGN. 
In the optical the conventional dividing line between Seyfert galaxies
and QSOs is $M_{\rm B}=-23$. Despite their high intrinsic luminosities,
type II QSOs are very hard to identify. A significant fraction of the emitted
power is absorbed by the optically thick torus. Furthermore, their lack of
emission lines in a wide optical wavelength range hampers their
identification. About 150 
optically selected type II QSOs from the Sloan Digital Sky Survey 
were studied in \cite{zakamska}. However type II QSOs are more
efficiently found
in follow-up observations of X-ray surveys 
(\citealt{szokoly}; \citealt{mainieri}; \citealt{krumpe}; \citealt{barcons2007}; 
\citealt{tedds}). 

Type II QSOs play an important role in understanding the X-ray universe.
Since they show significant absorption, type II QSOs/AGN are considered to 
be a main contributor to the hard X-ray background and their existence in 
considerable numbers is needed for the synthesis of the X-ray background
(\citealt{gilli}). 
Possible evolution of the absorption with intrinsic luminosity and/or redshift
is a matter of intensive debate and would essentially influence the X-ray background
synthesis models.

Many papers discuss the fraction of absorbed AGN as a function of luminosity
and/or redshift, and this is a highly controversial subject 
(\citealt{dwelly}; \citealt{tozzi}; \citealt{akylas}). 
However, there are only a few studies (e.g. \citealt{lafranca}) which
investigate the evolution of absorbing column 
densities in type II AGN up to high redshift. Other studies 
focus on the local universe, where the existence of so-called 
Compton-thick absorbed objects is well known 
(\citealt{maiolino}; \citealt{bassani}; \citealt{risaliti}). 
A Compton-thick object is absorbed by
column densities of $N_{\rm  H} > 1.5 \times 10^{24}$\,cm$^{-2}$, so that
the cross section of Compton scattering overcomes the photo-electric
absorption. Hence, the reflected X-ray component is observable.

There is evidence for a large fraction of Compton-thick absorbed quasars at high
redshifts (\citealt{martinez-sansigre}), but only one candidate has been
reported so far at high redshift (\citealt{norman}), and even in this case the
column density is quite uncertain. 
Most of the X-ray sources from deep {\it Chandra} and {\it XMM-Newton} 
surveys are absorbed by column densities $10^{21} < N_{\rm  H}/{\rm cm^{-2}} < 10^{23}$,
with hard X-ray luminosities of $10^{42} < L_{\rm X_{\rm OBS}}/({\rm erg/s}) < 10^{44}$ 
(\citealt{comastri}; \citealt{szokoly}; \citealt{mainieri}; \citealt{mateos}). These values are
also found in Compton-thin absorbed ($10^{21} < N_{\rm  H}/{\rm cm^{-2}} < 10^{24}$) 
Seyfert II galaxies at low redshifts.

The definition of type II QSOs is somewhat arbitrary.
\cite{zakamska} define a type II QSO only based on its optical properties. 
They select AGN with narrow permitted emission lines (FWHM $<2000$\,km/s) and 
classify objects with [\ion{O}{iii}] $\lambda$5008 line luminosity of 
$L \geq 3 \times 10^8\,L_{\odot}$ as type II QSO. 
\cite{mainieri} introduce the term ``type II QSO region''. Although 
their type I/II classification is based on optical spectra, a type II QSO
has to have an intrinsic X-ray luminosity of $L_{\rm X_{\rm INT}}>10^{44}$\,erg/s 
(0.5-10\,keV) and an absorbing hydrogen column density of 
$N_{\rm H}>10^{22}$\,cm$^{-2}$. Throughout the paper we use the definition of
a type II QSO to be an object with narrow forbidden and narrow permitted
emission lines in the optical spectrum, as well as a de-absorbed intrinsic 
X-ray luminosity $L_{\rm{X_{INT}}}>10^{44}$\,erg/s 
(0.5-10\,keV band). However, the Seyfert type II classification of our sample
objects is strictly based on the optical spectra.

The present work focuses on properties of the high X-ray luminosity type II
QSOs found in medium deep surveys performed by us. 
The paper is organised as follows. In Sect.~\ref{sample} we describe
how we selected our type II QSO candidate sample and we study the properties 
in Sect.~\ref{properties}. The X-ray data and the extraction of the X-ray
spectra are described in Sect.~\ref{xrayobservation}. Since X-ray spectra 
with small numbers of counts have to be fitted, we studied different fitting and binning 
methods in Sect.~\ref{xrayanalysis} and then describe the \mbox{analysis} of the X-ray
spectra. The results are discussed in 
Sect.~\ref{discussion}. Finally, our conclusions are outlined in
Sect.~\ref{conclusions}.

Unless mentioned otherwise, all errors refer to a $68\%$ confidence interval.
We assume $H_0=70$\,km/s/Mpc, $\Omega_{\rm M}$=0.3, and
$\Omega_{\Lambda}$=0.7. 

\section{Definition of the type II QSO sample \label{sample}}
We began with a list of 51 spectroscopically classified type II QSO
candidates which are associated with X-ray sources. All optical counterparts
have $z \ge 0.5$ and observed X-ray luminosities (not corrected for intrinsic
absorption) of log($L_{\rm X_{OBS}}$)$ > 42.5$. 
The type II QSO candidates were taken from three different X-ray surveys, the
{\it XMM-Newton} Marano field survey (\citealt{krumpe}), the {\it XMM}-2dF wide angle
survey (XWAS, \citealt{tedds}), and from 'An {\it XMM-Newton} International Survey' 
(AXIS, \citealt{barcons, barcons2007}; \citealt{carrera}). 
To obtain a well-defined sample with secure optical type II identifications and
X-ray counterparts we applied a two stage process.
First, we visually inspected the optical spectra of all type II QSO candidates
from all X-ray surveys and determined the FWHM of the emission lines. 
The type II QSO candidate classification was only based on the optical spectra. 
The spectral resolution of the optical setup was $\sim$1050\,km/s for the Marano
field survey, $\sim$700\,km/s for XWAS, and $\sim$300-600\,km/s for AXIS. 
Therefore, an intrinsic line of 1200\,km/s was observed with
$\sim$1250-1600\,km/s. We defined emission lines with a measured resolution 
(including instrumental resolution) of FWHM $\le1500$\,km/s 
as narrow.

The optical spectra were only used to determine a secure Seyfert type of the 
objects. They were not used to separate between AGN and QSOs. None of the 
51 type II QSO candidates showed obvious broad emission lines in the optical spectra. 
However, since we aimed for a very strict type II QSO sample, we excluded
many objects that showed only the low excitation \ion{O}{ii} line.
The optical spectrum had to comprise several emission lines that allowed us to
establish an optically secure classification of the Seyfert type. In addition
the signal-to-noise ratio had to be appropriate in order to verify the
non-existence of broad permitted emission line. We excluded 
all doubtful cases.       

We introduced two different categories of type II 
QSO candidates based on the reliability of the type II classification. 
Objects that have optical spectra with high excitation lines and at least one 
permitted AGN emission line that is detected but narrow
(e.g. \ion{Ly}-$\alpha$, \ion{C}{iv}, H$\beta$) are marked with the optical
flag 'S' (see Table~\ref{table1}) - signifying a secure optical identification. 
Their optical classification is very robust.

Objects with less secure type II identification belong to the tentative
sample (optical flag 'T'). The classification as tentative object can be due to
the following reasons: 
\begin{itemize}
 \item{The optical spectrum shows narrow emission lines, common AGN high 
       excitation lines, but no \ion{Mg}{ii} emission. 
       Only [\ion{Ne}{v}] and [\ion{O}{ii}] are detected in the 
       spectra.
       The presence of a common AGN high excitation line ([\ion{Ne}{v}]) 
       without a broad \ion{Mg}{ii} suggests the classification as a type II QSO candidate. 
       However, since no other permitted emission line is  
       accessible, a secure type II classification cannot be established.}
 \item{Common AGN high excitation lines are narrow but show 
       indications of underlying broad components (optical spectra in
       Appendix~\ref{appendix1}).
       These objects are likely to be transition objects between Seyfert type II / 
       type I}
 \item{The signal-to-noise ratio of the H$\beta$ line, or its coincidence
       with the atmospheric band prevents a robust determination of the FWHM
       and could hide a weak, but broad H$\beta$ component.}  
\end{itemize}
For all sources in the tentative sample we give an individual comment at the
end of this section. 
After inspection of the optical spectra we were left with 14 secure objects 
and 12 type II objects that belong to the tentative sample.

As a second step to setting up a well-defined type II QSO sample, 
we verified that the spectroscopically identified counterpart is
associated with the X-ray source. For a significant number of objects much 
deeper imaging data have become available subsequent to the epoch when the
spectra were obtained. For example, the XWAS optical counterparts were originally selected from 
SuperCosmos optical imaging survey data (\citealt{hambly}). 
For the majority of the objects we now have additional, deeper imaging data in different bands. 
These imaging data were obtained with the Wide Field Imager (WFI, 2.2\,m
telescope at La Silla) or the Wide Field Camera (WFC, 2.5\,m Isaac Newton telescope
on La Palma). 
We visually investigated the best available 
imaging data (WFI, WFC, SuperCosmos) for additional 
optical counterparts and rejected doubtful counterpart identifications. 
Furthermore, we computed the probability that the optical object with the 
$R$-band magnitude $m_{\rm{R}}$ is associated with the X-ray source. This was based on   
\begin{eqnarray}\label{equation1}
L=\frac{Q\,exp(-dist_{\rm{OX}}^2/2)}{2\pi \sigma_{\rm{X}}^2 N(<m_{\rm{R}})}
\end{eqnarray}
where $dist_{\rm OX}$ is the distance between optical counterpart and X-ray
source and $\sigma_{\rm{X}}$ is the position error of the X-ray source 
(\citealt{sutherland}). A detailed description is given in \cite{krumpe}.
Most of our optical counterparts have probabilities of $>95$\% that they are associated 
with the X-ray source. We rejected objects that have probabilities of less
than 65\%. All objects in the secure sample have probabilities $\ge$83$\%$.

After the X-ray counterpart verification our sample consists of 22 sources including 13
secure and 9 tentative objects. The separation between AGN
and QSOs was made after the determination of the de-absorbed intrinsic X-ray luminosity (see
Sect.~\ref{xrayanalysis}).\vspace{0.0cm}\\

Comments on the optical spectra of tentative objects:\\
{\bf Marano 32A} -- all narrow emission lines have underlying broad (mainly
blue-shifted) components.\\
{\bf Marano 47A} -- the signal-to-noise ratio of the optical spectrum does not allow the exclusion of the presence of a weak, broad H$\beta$ emission line.\\
{\bf Marano 50A} -- H$\beta$ and [\ion{O}{iii}] not covered by the spectrum,
however, most likely a type II object since no \ion{Mg}{ii} but [\ion{O}{ii}] and 
[\ion{Ne}{v}] emission lines.\\
{\bf Marano 51A} -- H$\beta$ and [\ion{O}{iii}] covered by the spectrum
but H$\beta$ coincides with the edge of the atmospheric A-Band. However, a strong broad
H$\beta$ emission line can be ruled out and [\ion{O}{ii}], 
[\ion{Ne}{v}] emission lines are visible.  \\
{\bf Marano 66A} -- SNR of the optical spectra does not 
exclude the presence of a weak, broad H$\beta$ emission line ([\ion{O}{ii}] and [\ion{Ne}{v}] emission).\\
{\bf Marano 116A} -- type II object/X-ray bright optical normal galaxy (XBONG)
- galaxy spectrum with prominent
[\ion{O}{ii}] emission line, no \ion{Mg}{ii}, very weak [\ion{O}{iii}]
emission lines, H$\beta$ coincides with the edge of the A-Band.\\
{\bf Marano 133A} -- some narrow emission lines have underlying broad,
blue-shifted components.\\
{\bf Marano 253A} -- type II object/XBONG - no \ion{Mg}{ii} and
H$\beta$ emission, weak [\ion{O}{iii}] emission, the second [\ion{O}{iii}] 
line falls into the atmospheric A-band.\\
{\bf sds1b-014} -- high SNR spectrum, H$\beta$ and [\ion{O}{iii}] not covered 
by the spectrum, however most likely a type II object since no \ion{Mg}{ii} 
but [\ion{O}{ii}] and [\ion{Ne}{v}] emission lines.\\

\section{Properties of the type II QSO candidate sample\label{properties}}
In Table~\ref{table1} we summarise the observed
properties of the objects. We list the name of the spectroscopically identified 
counterpart, optical 
coordinates, distance between
spectroscopically identified counterpart and X-ray source, the WFI-$R$-band
magnitude (unless otherwise mentioned), redshift, optical flag, X-ray
counterpart probability, count rate, 0.5-10\,keV flux, and the Galactic
absorption along the line of sight.
 
\begin{figure}
  \centering
  \resizebox{\hsize}{!}{ 
  \includegraphics[bbllx=95,bblly=365,bburx=550,bbury=712]{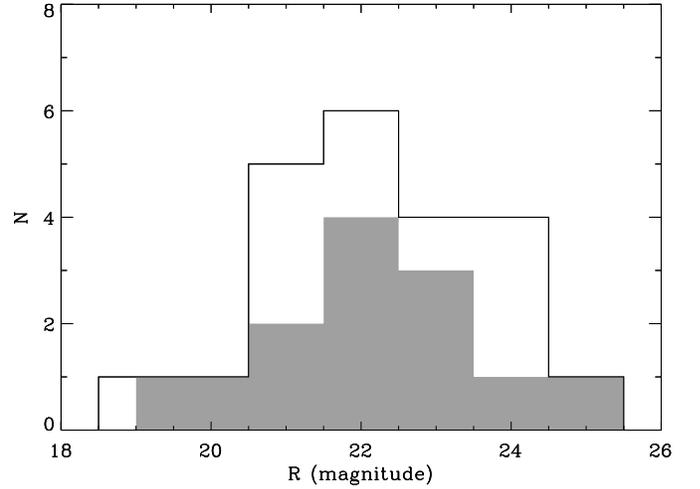}} 
      \caption{$R$-band magnitude histogram of all 22 type II QSO candidates
      (black solid line) and of the 13 secure type II QSO candidates (grey
      filled histogram).}
         \label{fig:R_histogram}
\end{figure}

The objects cover a redshift range of 
$0.5\le z \le 3.278$. Since the lack of emission lines in a wide optical
wavelength range hampers the identification of type II objects, 
we only have one type II QSO candidate in the redshift interval of $z=1-2$.
Figure~\ref{fig:R_histogram} shows the 
$R$-band magnitude histogram of the selected 
objects. As a simplification we consider the $r$-band (SDSS) magnitude to be
equal to the $R$-band magnitude, although we are aware that shifts of 0.5 
magnitude between $R$ and $r$ (SDSS) may occur. 
The majority of our sources have $21<R<24$. 
\begin{figure}
  \centering
  \resizebox{\hsize}{!}{ 
  \includegraphics[bbllx=55,bblly=360,bburx=560,bbury=715]{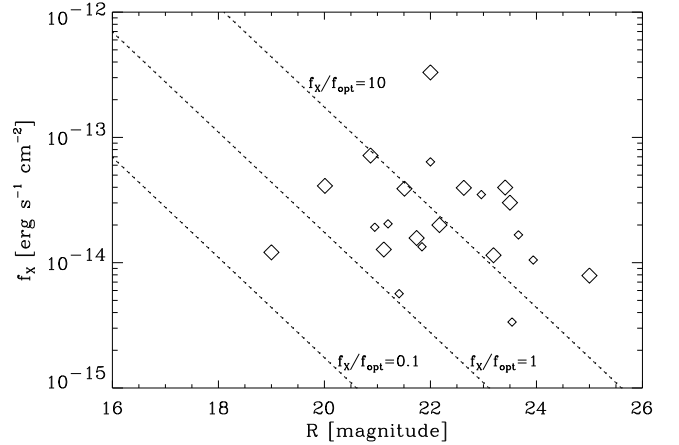}} 
      \caption{Observed 0.5-10\,keV X-ray flux vs. $R$-band magnitude. 
               Large symbols represent secure 
               type II QSO candidates while small symbols illustrate 
               tentative type II QSO candidates.
               Dashed lines indicate $f_{\rm X}/f_{\rm OPT}$ values of
               0.1, 1, and 10.}
         \label{fig:R_fx}
\end{figure}
The \mbox{0.5-10\,keV} flux in the sample ranges from $0.3-33\times10^{-14}$\,erg/s/cm$^2$.

To calculate $f_{\rm X}/f_{\rm OPT}$ values we derived the optical fluxes 
in a band centred at 7000\,\AA\ with a width of 1000\,\AA\ using the
equation $f_{\rm{opt}}=10^{-0.4 R - 5.759}$ (\citealt{zombeck}).
X-ray fluxes were calculated for the 0.5-10\,keV energy range. 
The $R$-band magnitude vs. X-ray flux plane (Figure~\ref{fig:R_fx}) 
clearly shows AGN activity in all selected sources, since they 
have X-ray-to-optical flux ratios of $f_{\rm X}/f_{\rm OPT} > 0.25$. 
In accordance with \cite{mainieri} we consider objects with 
$f_{\rm X}/f_{\rm OPT} > 0.1$ as AGN. 
Almost half of the selected type II QSO candidates have noticeable 
high X-ray--to--optical flux ratios ($f_{\rm X}/f_{\rm OPT}>10$). 
\cite{szokoly} mentioned that type II AGN/QSOs cluster at higher
X-ray--to--optical flux ratios than type I AGN. The majority of the 
spectroscopically classified type II AGN/QSOs in the Lockman Hole 
also show high $f_{\rm X}/f_{\rm OPT}$ values (\citealt{mainieri}). 
The observed 0.5-10 keV X-ray luminosities (not corrected for intrinsic absorption) 
of our objects range from $L_{\rm X_{\rm OBS}}\sim 10^{43-45}$\,erg/s.

\begin{onecolumn} 
\begin{table}
\caption[]{Observed properties of type II QSO sample} 
\begin{center}\label{table1}
 \begin{tabular}{rrlclcclccc}\hline
   (1)&(2)&(3)&(4)&(5)&(6)&(7)&(8)&(9)&(10)&(11)\\
$No$&$RA$&$DEC$&$r_{\rm{OX}}$&$R$&$z$&Opt.&XID    &CR&Flux&$N_{\rm H_{GAL}}$\\
    &    &     &             &   &   &flag  &probab.&  &$10^{-14}$& $10^{22}$\\
    &    &     &[arcsec]        &   &  & & &[$ks^{-1}$]&[${\rm erg\,s^{-1}\,cm^{-2}}$]& [cm$^{-2}$]\\\hline
  Marano 9A &  3 15 09.9 & -55 13 13 & 1.18 & 22.63         & 1.427 & secure     & 0.99         & 21.3$\pm$0.9 & 3.957 & 0.027 \\ 
 Marano 20A &  3 16 21.5 & -55 17 59 & 0.63 & 23.5$^E$      & 2.207 & secure     & 0.97$^K$     & 17.1$\pm$1.3 & 3.014 & 0.027 \\ 
 Marano 32A &  3 15 47.0 & -55 17 55 & 1.44 & 22.96         & 2.727 & tentative     & 0.99         & 12.5$\pm$0.9 & 3.497 & 0.027 \\ 
 Marano 39A &  3 13 39.7 & -55 01 51 & 0.95 & 23.41         & 0.862 & secure     & 0.98         & 19.8$\pm$1.5 & 3.982 & 0.027 \\ 
 Marano 47A &  3 15 38.7 & -55 10 44 & 1.34 & 23.94         & 0.900 & tentative     & 0.95         &  4.5$\pm$0.5 & 1.051 & 0.027 \\ 
 Marano 50A &  3 14 09.9 & -55 17 46 & 1.25 & 23.66         & 0.986 & tentative     & 0.96         &  7.6$\pm$0.7 & 1.667 & 0.027 \\ 
 Marano 51A &  3 16 30.6 & -55 15 03 & 1.83 & 20.95         & 0.58  & tentative     & 0.98         & 10.4$\pm$1.0 & 1.919 & 0.027 \\ 
 Marano 63A &  3 15 16.9 & -55 06 02 & 1.25 & 23.19         & 2.800 & secure     & 0.97         &  7.2$\pm$0.7 & 1.147 & 0.027 \\ 
 Marano 66A &  3 15 00.7 & -55 07 18 & 1.23 & 21.41         & 0.981 & tentative     & 0.99         &  4.0$\pm$0.6 & 0.566 & 0.027 \\
Marano 116A &  3 16 20.7 & -55 16 52 & 1.30 & 21.2$^E$      & 0.581 & tentative     & 0.99$^K$     &  5.4$\pm$0.9 & 2.043 & 0.027 \\
Marano 133A &  3 14 26.2 & -55 21 13 & 1.02 & 23.54         & 2.321 & tentative     & 0.92         &  2.7$\pm$0.6 & 0.336 & 0.027 \\
Marano 171A &  3 13 51.4 & -55 02 56 & 2.18 & 22.17         & 0.800 & secure     & 0.86         &  5.8$\pm$0.8 & 1.998 & 0.027 \\
Marano 224B &  3 13 04.8 & -55 16 04 & 1.70 & 21.51         & 0.690 & secure     & 0.96         &  8.9$\pm$1.2 & 3.899 & 0.027 \\
Marano 253A &  3 14 38.0 & -55 06 50 & 2.64 & 21.84         & 0.517 & tentative     & 0.65         &  2.6$\pm$0.5 & 1.341 & 0.027 \\
Marano 463A &  3 16 25.3 & -55 08 39 & 0.55 & 25.0$^E$      & 2.531 & secure     & 0.98$^K$     &  3.1$\pm$0.8 & 0.789 & 0.027 \\
Marano 610A &  3 15 51.8 & -55 12 22 & 1.25 & 21.12         & 0.699 & secure     & 0.98         &  3.6$\pm$0.6 & 1.275 & 0.027 \\   
X21516\_135 &  2 26 26.7 & -04 36 56 & 0.90 & 21.74$^r$     & 3.278 & secure     & 0.91         &  0.9$\pm$0.2 & 1.574 & 0.025 \\
X00851\_154 & 22 15 41.6 & -17 37 54 & 0.64 & 19.00         & 2.976 & secure     & 0.96         &  1.7$\pm$0.2 & 1.210 & 0.023 \\
X01135\_126 &  1 52 57.6 & -14 08 40 & 0.91 & 20.87$^{UK}$  & 0.543 & secure     & 0.97$^{WFC}$ & 10.5$\pm$0.5 & 7.167 & 0.015 \\
X03246\_092 &  0 43 46.0 & -20 29 56 & 2.31 & 20.01$^{UK}$  & 0.500 & secure     & 0.83$^{WFC}$ &  3.2$\pm$0.8 & 4.104 & 0.015 \\
phl5200-001 & 22 28 26.4 & -05 18 20 & 0.63 & 22.0$^r$      & 0.711 & secure     & 0.99$^{WFC}$ & 50.0$\pm$1.2 & 33.03 & 0.053 \\
sds1b-014   &  2 18 43.0 & -05 04 37 & 0.80 & 22.0$^r$      & 0.962 & tentative     & 0.98$^{WFC}$ &  6.1$\pm$0.4 & 6.380 & 0.026 \\\hline
\end{tabular}
\end{center} 
\end{table}
{\tiny
Comments for Table~\ref{table1}:\\
E -- FORS $R$ pre-imaging magnitude estimate; cross-calibrated with WFI-$R$-band
magnitude catalogue; magnitude error estimate $\pm$0.5\,mag.\\
K -- since no $R$-band magnitude catalogue entry existed, the X-ray-to-optical counterpart 
probability calculation (see Eq.~\ref{equation1}) is based on the 
$K$-band magnitude catalogue.\\
r -- SDSS $r$-band magnitude (AB magnitude).\\
UK -- UK-Schmidt red plate magnitude.\\
WFC -- since no $R$-band magnitude catalogue entry existed, the X-ray-to-optical counterpart 
probability calculation (see Eq.~\ref{equation1}) is based on the WFC-data.\\
}
\end{onecolumn}
\begin{twocolumn}


\section{XMM-Newton observations and extraction of X-ray spectra\label{xrayobservation}}

All data were processed with the {\tt SAS} version 7.0
(Science Analysis Software, \citealt{gabriel}) package and corresponding calibration
files. The {\tt epchain} and {\tt emchain} tasks were used for generating
linearised event lists from the raw PN and MOS data. 
For all sources the effects of photon pile-up was negligible.

The {\it XMM-Newton} data reduction for the Marano field survey is described 
in detail in \cite{krumpe}. For the XWAS and AXIS sources we downloaded 
all available X-ray data from the {\it XMM-Newton}
archive\footnote{http://xmm.esac.esa.int/xsa} 
up to and including July 2007. Periods of high background were excluded from the analysis 
of all relevant data sets in the standard way. 

Circular or box-shaped source and background regions were manually 
determined for all 
contributing observations. Sources at large off-axis angles in
the contributing observations were not considered as follows. The largest offaxis angles were 
720 arcsec for the PN and 820 arcsec for the MOS detectors, respectively.
The lower PN area was a result of the enhanced background contamination near
the edges of the PN detector.
Additional X-ray sources in the background regions were masked out. The 
auxiliary response file (arf) was computed for each source and observation 
individually. 
For the Marano field sources, we used appropriately ``canned'' response 
matrices from the {\it XMM-Newton} calibration 
homepage\footnote{http://xmm.esac.esa.int/external/xmm\_sw\_cal/calib} 
({\it XMM}-revolution 110 and pattern 0-12).
The PN y-coordinate of an X-ray source was determined to link the relevant 
PN response matrix file (rmf for single and double
events, version 6.8) to the X-ray spectrum of the source.
For the XWAS and AXIS sources the response matrix files were computed
on a case-by-case basis. Where multiple X-ray spectra of a given source were added, 
the mean rmf was computed as a weighted mean. For details of the procedure 
see \cite{page}. The two MOS spectra were always added to form a single MOS 
spectrum. 

The X-ray source of the secure type II object phl5200-001 is surrounded 
by diffuse X-ray emission. For the reduction of the X-ray spectrum we 
used a smaller extraction radius for the point source and the diffuse X-ray 
emission as the background region.

\section{X-ray spectral analysis\label{xrayanalysis}}
The X-ray spectral analysis was performed with {\tt XSPEC} (\citealt{arnaud}) version 12.3.0.
Although we have a few objects with several hundreds of net PN source counts in the 0.2-8 keV range, 
the distribution peaks at $\sim$40 net PN source counts. An appropriate X-ray spectral 
analysis for the low count regime has to be found. Although we 
only refer to the net PN counts in our simulations and in 
Table~\ref{table2}, the fit uses both the PN and MOS data. 
Hence, the total number of counts used by the fit is typically twice that given in 
Table~\ref{table2}.

\subsection{Defining the appropriate fit statistic and binning method\label{fitmethod}}
A problem notorious to X-ray astronomy is proper fitting of spectra with relatively few counts.
\cite{tozzi} approached this problem for the X-ray sources in the {\it Chandra}
Deep Field South by running simulations for two different fitting
procedures: Cash-statistics (unbinned) and classic $\chi^2$-statistics with a
binning of 10 counts per bin (min 10). They concluded that the unbinned 
Cash-statistic fits recovered the input values better for X-ray spectra 
with less than 50 counts.

We carried out a much more extensive study of the fit results as a function
of fit statistic and binning method. This investigation can 
be used to study the error distribution of the intrinsic 
column density ($N_{\rm H}$) and to determine how many PN net counts 
are required to perform free fits of $N_{\rm H}$ and photon index $\Gamma$ 
with acceptable errors in both parameters.

We assumed emission from the AGN to be described as a power law with photon
index $\Gamma$, modulated by Galactic foreground absorption and further
intrinsic cold absorption at the redshift of the AGN.  
Following \cite{mainieri} and \cite{mateos}, we simulated X-ray spectra with 
an input value of $\Gamma$ = 2. All simulated X-ray spectra included Galactic absorption of 
$N_{\rm H_{GAL}}=2\times10^{20}$\,cm$^{-2}$ and considered all 
possible parameter combinations of Table~\ref{simulationtable1}.

The X-ray spectra were normalised to reach the desired \mbox{0.2-8\,keV} PN net counts
with a deviation of up to 5\%. We added Poisson noise to the X-ray spectra.
We used a typical representative background file (Marano 9A). 
As for the real data we added MOS1 and MOS2 spectra 
to form a single MOS spectrum. The same source was simulated 1000 times for each set
of specific parameters.

The simulated X-ray spectra were grouped by using different binning methods 
(Table~\ref{simulationtable2}). Cash and $\chi^2$-statistic were applied to 
recover the $N_{\rm H}$ input value. The redshift and the Galactic absorption 
were set to the input value. The nominal initial guess for the intrinsic
absorption was $N_{\rm H}=0$\,cm$^{-2}$, but other initial guesses up to 
$N_{\rm H}= 10^{24}$\,cm$^{-2}$ were tested as well.

\begin{table} 
\begin{center} 
\caption{Set of input parameters for the simulated X-ray spectra.}\label{simulationtable1}  
 \begin{tabular}{rl} \hline 
   Parameter                 & value \\\hline
   0.2-8\,keV net counts     & 10, 40, 100, 130, 200\\
   Redshift                  & 1, 2, 3\\
   Column density [cm$^{-2}$]& 0, 10$^{21}$, 10$^{22}$, 10$^{23}$, 10$^{24}$\\
   Photon index $\Gamma$     & 2.0\\\hline
 \end{tabular} 
\end{center}  
\end{table} 
\begin{table} 
\begin{center} 
\caption{Fit statistics and binning methods for the simulated X-ray
         spectra$^{3}$.}\label{simulationtable2}  
 \begin{tabular}{rl} \hline 
   Fit statistic         & binning methods \\\hline
   Cash-statistic        & unbinned, min 1, min 2, min 3,\\ 
                         & min 5, min 10, min 15, channel 60,\\
                         & channel 120, channel 180, channel 240\\
   $\chi^2$-statistic    & min 10, min 15 \\
   Photon index $\Gamma$ & free fit, 2.0\\\hline
 \end{tabular} 
\end{center}  
\end{table} 
\footnotetext[3]{The data were grouped from a minimum PHA channel 
         (corresponding to 0.2\,keV) to the maximum PHA channel 
         (corresponding to 8\,keV), with at least $n$ counts per bin 
         (abbreviated as min $n$) or with a fixed number $n$ of channels
         (channel $n$).}
We studied the recovered $N_{\rm H}$ distribution with different 
fit statistics and binning methods by analysing the number of mismatches, the
peak, the shape and the significance of $N_{\rm H}$ detection. The best fit statistic and binning method
was selected by these criteria in the given order. In summary, the
most important results of our simulations are as follows. 
\begin{itemize}
  \item{With only very few exceptions, the Cash-statistic with a binning of one
        count per bin (min 1) recovered the input values best.}
    \begin{itemize} 
      \item{Even at a level of 40 net PN counts, the Cash-statistic with a
            binning of minimum one count per bin correctly recovers more than 
            90\% of the objects with an intrinsic absorption of 
            $N_{\rm H}\sim 10^{22}-10^{24}$\,cm$^{-2}$ 
            up to $z=3.5$. Figure~\ref{fig:Nh1_z2_40.ps} shows the retrieved 
            $N_{\rm H}$ distribution for an input value of 
            $N_{\rm H} = 10^{22}$\,cm$^{-2}$ at $z=2$.
            The variance of the recovered $N_{\rm H}$ distribution 
            increases with redshift. However, individual peaks clearly 
            separate in the studied redshift range, when the intrinsic 
            $N_{\rm H}$ values differ by a factor of approximately 10 in 
            hydrogen column density. The recovered $N_{\rm H}$ distribution 
            for an intrinsically unabsorbed source at $z=1$ is shown in 
            Fig.~\ref{fig:Nh0_z1_40.ps}. Between 50-60\% of all objects are 
            correctly recovered independent of redshift. 
            Although not all fits converge at $N_{\rm H} = 0$\,cm$^{-2}$, 
            none of the resulting $N_{\rm H}$ values is significantly different
            from zero. The peak of the
            misclassified absorbed objects shifts from 
            $N_{\rm H}\sim 10^{21}$\,cm$^{-2}$ for $z=1$ to 
            $N_{\rm H}\sim 7\times 10^{21}$\,cm$^{-2}$ for $z=3$.}
      \item{For 10 net PN counts the fit retrieves a very broad 
            $N_{\rm H}$ distribution for objects with an intrinsic absorption of 
            $N_{\rm H}\sim 10^{22}-10^{24}$\,cm$^{-2}$. Independent of
            redshift the fit recovers roughly 30\% of the intrinsically 
            absorbed X-ray sources as unabsorbed sources 
            ($N_{\rm H}=0$\,cm$^{-2}$). However, the fit does not significantly overpredict 
            the $N_{\rm H}$ values. Although no absorption 
            is recovered with a significance above 2$\sigma$ due to the large errors, the fitted 
            $N_{\rm H}$ values can be used as an estimate for the intrinsic
            $N_{\rm H}$ values. An unabsorbed X-ray source is correctly
            recovered by the fit in 50-60\% of all cases. In 90\% of the cases
            the fitted             
            $N_{\rm H}$ values for an intrinsically unabsorbed X-ray 
            source do not exceed $N_{\rm H}\sim 10^{22}$\,cm$^{-2}$ for $z=1$
            and $N_{\rm H}\sim 3 \times 10^{22}$\,cm$^{-2}$ for $z=2$.}
    \end{itemize}
  \item{The unbinned Cash-statistic as used in \cite{tozzi} 
        shows the narrowest distribution in $N_{\rm H}$ 
        for most of the parameter combinations, but the peak of the recovered
        distribution was found to be strongly dependent on 
        the initial guess of $N_{\rm H}$ and weakly favours 
        $N_{\rm H}=10^{22}$\,cm$^{-2}$ independent of input parameter for 
        net PN counts less than 100.}
 \item{For a binning of min 10 or min 15 we find no difference between the Cash and 
       the $\chi ^2$-statistic even at a level of 40 net PN counts. Only the significance of
       the absorption is marginally higher for the Cash-statistic. There is also no difference
       between a binning of min 10 and min 15.} 
 \item{Free fits in $N_{\rm H}$ and $\Gamma$ are acceptable above 100 PN net
       counts. The recovered $N_{\rm H}$-values show a narrow 
       distribution 
       at the input $N_{\rm H}$ value (FWHM$\sim$0.4\,dex). For all 
       combinations of fitting and binning methods we retrieve the peak 
       of $\Gamma \simeq 2$. For a binning of minimum one count per bin (min 1)
       the recovered $\Gamma$ distribution clearly peaks at $\Gamma$=2 with FWHM$\sim$0.8 
       (FWHM$_{\rm 130 counts}\sim$0.4, FWHM$_{\rm 200 counts}\sim$0.2).}
\end{itemize}
\begin{figure}
  \centering
  \resizebox{\hsize}{!}{ 
  \includegraphics[bbllx=80,bblly=30,bburx=560,bbury=340]{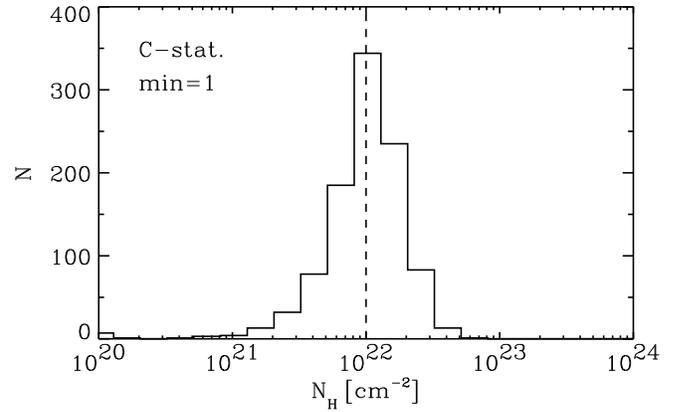}} 
      \caption{Recovered $N_{\rm H}$ distribution for an X-ray source with 40 net PN counts at 
                                 $z=2$. The input $N_{\rm H}=10^{22}$\,cm$^{-2}$ is 
               indicated by the dashed line. Fit method used: fixed $\Gamma =
               2$, Cash-statistic,
               binning: min 1; 1000 simulations. 
              }
         \label{fig:Nh1_z2_40.ps}
\end{figure}
\begin{figure}
  \centering
  \resizebox{\hsize}{!}{ 
  \includegraphics[bbllx=80,bblly=30,bburx=560,bbury=340]{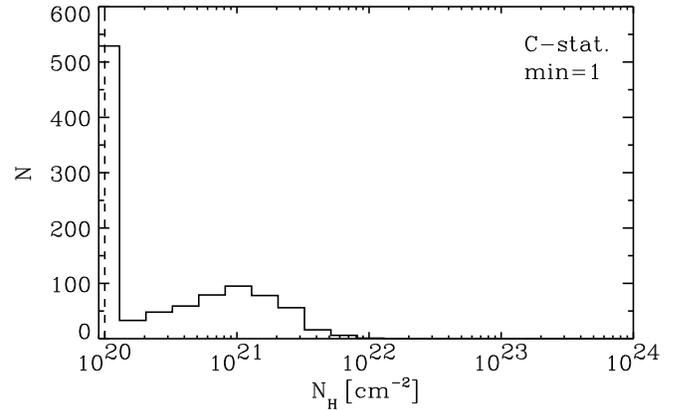}} 
      \caption{Recovered $N_{\rm H}$ distribution of an unabsorbed X-ray source with 40 net PN  
               counts at $z=1$. The input $N_{\rm H}=0$\,cm$^{-2}$ 
               is indicated by the dashed line. $N_{\rm H}=0$\,cm$^{-2}$ 
               is represented by the lowest bin. Fit method used: fixed $\Gamma =
               2$, Cash-statistic, binning: min 1; 1000 simulations. 
              }
         \label{fig:Nh0_z1_40.ps}
\end{figure}

Simulations as shown in Fig.~\ref{fig:Nh1_z2_40.ps} and 
Fig.~\ref{fig:Nh0_z1_40.ps} are used to determine the $N_{\rm H}$ range
that belongs to 68\% and 90\% of all simulations. In 
Fig.~\ref{fig:diagnostic} we show these distributions 
for simulations at different redshifts ($z$=0,1,2,3,3.5) and column densities
($N_{\rm H}=0, 10^{22}, 10^{23} $\,cm$^{-2}$). The simulations contain 
40 net PN counts and so are representative of many of our sources. 
The Cash-statistic with a binning of at least one count per bin is an appropriate method to 
determine the column density over a wide range of redshifts. It
does not significantly overestimate the input $N_{\rm H}$ values.
The fit recovers more than 90\% of the unabsorbed sources up to $z=3$ with 
$N_{\rm H}$ values below $N_{\rm H}= 10^{22}$\,cm$^{-2}$.

\begin{figure}
  \centering
  \resizebox{\hsize}{!}{ 
  \includegraphics[bbllx=65,bblly=360,bburx=560,bbury=715]{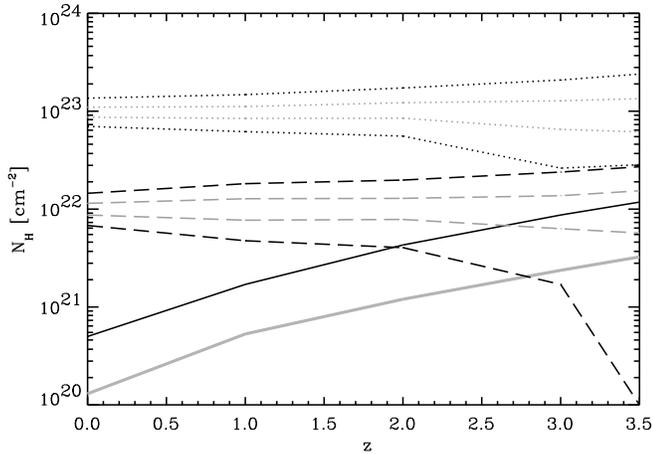}} 
      \caption{Diagnostic $N_{\rm H}$ vs. redshift plot. The
  plotted contours contain 68\% (grey) and 90\% (black) of the 
  simulated X-ray sources with 40 net PN 
  counts in the 0.2-8\,keV band (NOTE: The fits used more than twice as many
  counts since both the PN and MOS
  data are used). The solid lines are the limits for
  intrinsically unabsorbed sources. The limits for an intrinsic absorption of 
  $N_{\rm H}=10^{22}$\,cm$^{-2}$ are shown by dashed lines while dotted 
  lines represent $N_{\rm H}=10^{23}$\,cm$^{-2}$.}
         \label{fig:diagnostic}
\end{figure}

\subsection{Modelling the X-ray spectra}
Following our analysis in Sect.~\ref{fitmethod} we grouped the extracted
X-ray spectra in bins of at least one count per bin (0.2-8\,keV) and 
used the Cash-statistic to 
determine $N_{\rm H}$. The initial guess was set to $N_{\rm H}=0$.
We corrected for Galactic absorption and performed free fits in 
$N_{\rm H}$ and $\Gamma$ for spectra with more than 100 net PN 
counts. For X-ray spectra with less than 100 net PN counts 
we fixed $\Gamma=2$ and only fit $N_{\rm H}$. 
In these cases we give two errors for the value of $N_{\rm H}$ 
in Table~\ref{table2}. The first is the 1$\sigma$ error of $N_{\rm H}$ 
based on the fit with fixed $\Gamma=2$, 
while the second takes into account the systematic shift for 
different photon indices (deviation in $N_{\rm H}$ for a fixed 
$\Gamma=1.7$ and $\Gamma=2.3$). 
The observed (not corrected for intrinsic absorption) and the intrinsic 
0.5-10\,keV X-ray luminosities were computed by either a free fit in 
$N_{\rm H}$ and $\Gamma$ for the bright sources (net PN counts $>$100) or 
a fixed $\Gamma=2$ fit for the faint sources.

Due to contamination by soft detector background we only used the 0.3-8\,keV
band for the X-ray spectral analysis of X-ray source X03246\_092 (24 net PN 
counts in 0.3-8\,keV band). The soft energies in the X-ray spectra of Marano 
51A and Marano 610A could not be well fitted by an single absorbed power law. 
In these cases a fit of an absorbed power law plus a soft excess component 
reproduce the X-ray data better. Therefore, we used a two component fit, 
an absorbed power law and an unabsorbed power law. 
No evidence for soft excess was found in the X-ray spectra of the brightest
objects. The best fit models of the X-ray data are shown in 
Appendix~\ref{appendix1}.

Table~\ref{table2} shows the spectral parameters. In the columns 
we list the name of the spectroscopically identified counterpart (1), 
the XMMU source name (2) of the {\it XMM-Newton} X-ray source, the 
observed (3) and intrinsic (4) X-ray luminosity, the PN count number (5) in the 0.2-8\,keV band,
the absorbing hydrogen column density (6), the photon index (7), and the quality 
for the used fit (8) when the X-ray data are modelled by a power law plus 
intrinsic absorption model, and finally, the photon index (9) and the quality for
a purely reflection-dominated model (10). If no error is given for the photon
index, this parameter was fixed for the fitting.

\end{twocolumn}
\begin{onecolumn}
\begin{table}
\caption[]{Computed properties of type II QSO sample} 
\begin{center}\label{table2}
 \begin{tabular}{rccccccccc}\hline
   (1)&(2)&(3)&(4)&(5)&(6)&(7)&(8)&(9)&(10)\\
$No$&XMMU J&log$(L_{\rm{X_{OBS}}}$)&log$(L_{\rm{X_{INT}}}$)&$N_{\rm{Counts}}$&$N_{\rm H}$&$\Gamma$ &C-stat/d.o.f&$\Gamma$ &C-stat/d.o.f \\
& &0.5-10\,keV& 0.5-10\,keV &PN   & $10^{22}$&intr.& intr.&pexrav &pexrav \\
& & & & & [cm$^{-2}$]&absor.        & absor. & &      \\\hline
  Marano 9A&031510.1-551313 & 44.50 & 44.55 & 270&    0.3$\pm$0.1          &
  1.7$\pm$0.1 &   426/503 & 2.2$\pm$0.3 & 473/503   \\ 
 Marano 20A&031621.6-551759 & 44.82 & 44.99 &  64&    1.7$\pm$0.4$\pm$0.7  &
 2.0         &   183/227 &2.0 & 216/227     \\ 
 Marano 32A&031547.2-551755 & 44.88 & 45.22 & 129&    9.8$\pm$2.4          &
 1.8$\pm$0.2 &   285/291  & 2.3$\pm$0.6 & 295/291    \\ 
 Marano 39A&031339.7-550151 & 44.04 & 44.22 &  36&    0.8$\pm$0.2$\pm$0.4  &  2.0         &   111/117 &2.0&136/117     \\ 
 Marano 47A&031538.8-551043 & 43.49 & 43.72 &  38&    1.5$\pm$0.5$\pm$0.5  &  2.0         &   149/157 &2.0&180/157     \\ 
 Marano 50A&031410.1-551746 & 43.76 & 44.05 &  50&    2.5$\pm$0.6$\pm$0.8  &
 2.0         &   182/190 &2.0& 195/190      \\ 
 Marano 51A&031630.6-551501 & 43.37 & 43.61 &  47&    2.5$\pm$0.9$\pm$0.9$^*$  &  2.0         &   131/181  &2.0&132/183    \\ 
 Marano 63A&031517.1-550602 & 44.57 & 44.81 &  29&    4.1$\pm$1.4$\pm$1.4  &  2.0         &   125/140  &2.0&127/140    \\ 
 Marano 66A&031500.8-550718 & 43.36 & 43.45 &  39&    0.3$\pm$0.2$\pm$0.2  &  2.0         &   127/130 &2.0&129/130     \\
Marano 116A&031620.9-551651 & 43.31 & 43.72 &  28&    5.6$\pm$1.2$\pm$1.0  &  2.0         &   129/126  &2.0&136/126    \\
Marano 133A&031426.4-552113 & 44.04 & 44.04 &  14&    0.0$\pm$0.8$\pm$0.0  &  2.0         &    87/70   &2.0&94/70    \\
Marano 171A&031351.2-550257 & 43.61 & 44.00 &  14&    5.4$\pm$1.5$\pm$1.0  &  2.0         &    46/51  &2.0&63/51     \\
Marano 224B&031304.9-551606 & 43.71 & 44.25 &   8&   12.4$\pm$3.5$\pm$1.7  &  2.0         &    48/53   &2.0&70/53    \\
Marano 253A&031438.2-550648 & 42.97 & 43.55 &  11&   14.5$\pm$5.0$\pm$3.2  &  2.0         &    64/85  &2.0&73/85     \\
Marano 463A&031625.3-550839 & 44.37 & 44.53 &   7&    1.6$\pm$1.0$\pm$0.9  &  2.0         &    76/45  &2.0&81/45     \\
Marano 610A&031552.0-551222 & 43.28 & 43.75 &  17&   10.1$\pm$3.1$\pm$2.3$^*$  &  2.0         &   110/119 &2.0&113/120     \\   
X21516\_135&022626.7-043654 & 44.73 & 45.15 &  29&   14.8$\pm$5.7$\pm$4.8  &  2.0         &    61/79  &2.0&66/79     \\
X00851\_154&221541.6-173753 & 44.48 & 44.70 & 261&    5.6$\pm$2.0          &  1.6$\pm$0.1 &   572/621 &1.8$\pm$0.6&583/621     \\
X01135\_126&015257.5-140839 & 43.85 & 43.85 & 392&    0.0$\pm$0.1          &  1.7$\pm$0.1 &   355/513  &2.1$\pm$0.2&354/513     \\
X03246\_092&004345.8-202955 & 43.48 & 43.84 &  35&    3.8$\pm$1.2$\pm$1.3  &  2.0         &    74/67   &2.0&67/67      \\
phl5200-001&222826.4-051820 & 44.66 & 44.92 & 772&    4.5$\pm$0.4          &  1.6$\pm$0.1 &   834/970  &1.2$\pm$0.2&1200/970    \\
sds1b-014  &021842.9-050437 & 44.35 & 44.45 & 491&    0.4$\pm$0.1          &  1.8$\pm$0.1 &   546/593  &2.0$\pm$0.2&650/593    \\\hline
\end{tabular}
\end{center} 
\end{table}
{\tiny
Comments for Table~\ref{table2}:\\
$*$ -- a two component fit was used, an absorbed power law ($N_{\rm  H}$ is
given here) and an unabsorbed power law to model the soft excess
}
\end{onecolumn}

\begin{twocolumn}

\section{Discussion\label{discussion}}

In Fig.~\ref{fig:Nh_hist} we show the fitted intrinsic $N_{\rm H}$ 
distribution of the sources. A column density peak at 
$N_{\rm  H}=4\times10^{22}$\,cm$^{-2}$ is found. We find moderate 
absorption in the majority of our objects. The significance of the absorption
exceeds 2$\sigma$ in most of the cases (see the confidence contours in 
Appendix~\ref{appendix1}). Two type II objects are consistent with 
being unabsorbed X-ray sources, one of which is from the secure type II sample.

\begin{figure}
  \centering
  \resizebox{\hsize}{!}{ 
  \includegraphics[bbllx=95,bblly=364,bburx=560,bbury=705]{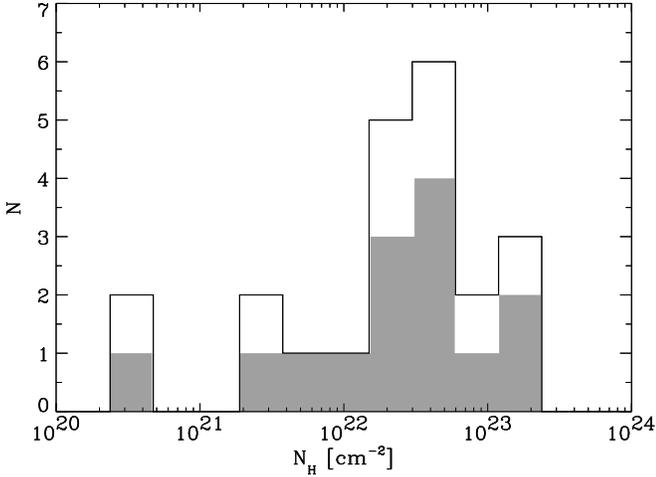}} 
      \caption{Intrinsic $N_{\rm H}$ distribution of all 22 type II objects 
               (black) and of the 14 type II QSOs (grey filled histogram).}
         \label{fig:Nh_hist}
\end{figure}

The determination of the intrinsic $N_{\rm H}$ also allows us to compute 
the de-absorbed intrinsic X-ray luminosity $L_{\rm{X_{INT}}}$ of our sources. 
As mentioned in Sect.~\ref{intro}, we define a type II object by the detection of 
narrow emission lines in the optical spectrum.  
Figure~\ref{fig:Nh_Lx} shows the intrinsic column density vs. 
de-absorbed intrinsic X-ray luminosity plane. The dividing line of 
\mbox{$L_{\rm{X_{INT}}} = 10^{44}$\,erg/s} is used to distinguish between type 
II AGN and QSOs.  
Ten type II objects from the secure sample and four tentative
objects are identified as QSOs. 
We detected one unabsorbed type II QSO but the object belongs to the
tentative sample. No obvious trend of absorption in type II QSOs 
with intrinsic X-ray luminosity is found. 

Most of the type II QSOs fall into the same region of the 
$N_{\rm H}$-$L_{\rm{X_{INT}}}$ diagram where previous studies have also found 
type II QSOs (\citealt{mainieri}; \citealt{szokoly}; \citealt{lafranca};
\citealt{ptak}). The additional criterion \mbox{($N_{\rm H}>10^{22}$\,cm$^{-2}$},
\citealt{mainieri}) makes only a small difference to our sample selection
(two more secure objects). 

\begin{figure}
  \centering
  \resizebox{\hsize}{!}{ 
  \includegraphics[bbllx=65,bblly=335,bburx=560,bbury=715]{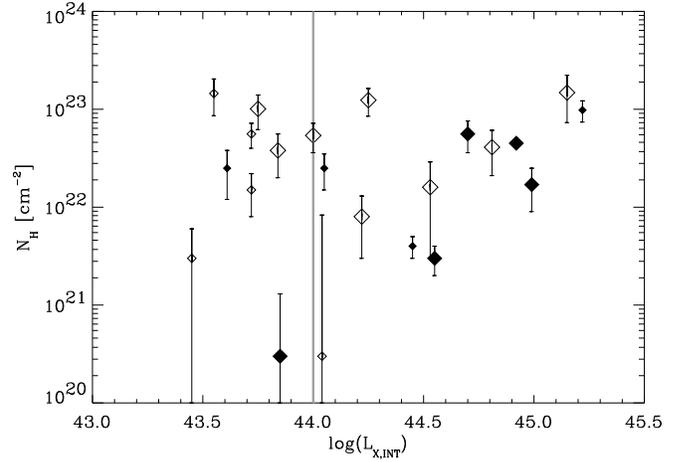}} 
      \caption{Intrinsic $N_{\rm H}$ vs. de-absorbed intrinsic 0.5-10\,keV X-ray
               luminosity. Large symbols represent optical secure 
               type II QSO candidates while small symbols illustrate 
               the tentative sample.
               Open symbols indicate X-ray sources that have less than 40 
               net PN counts in the \mbox{0.2-8\,keV} band; filled symbols $\ge$40 
               net PN counts. The vertical
               solid line at log$(L_{\rm{X_{INT}}}/(\rm{erg/s}))=44$ marks the dividing
               line between AGN and high luminosity QSOs. Objects with 
               log$(L_{\rm{X_{INT}}}/(\rm{erg/s}))\ge 44$ and 
               $N_{\rm  H}\ge 10^{22}$\,cm$^{-2}$ (upper right corner) 
               fall in the ``type II QSO region'' as defined by 
               \cite{mainieri}.}
         \label{fig:Nh_Lx}
\end{figure}

The $N_{\rm H}$-redshift plane for type II QSOs and type II AGN is 
shown in Fig.~\ref{fig:Nh_z}. We found no obvious differences between
the $N_{\rm H}$ distribution of AGN and QSOs. 
Although a tentative anticorrelation of $N_{\rm  H}$ vs. redshift below 
$z<1$ and a direct correlation above $z=1$ is indicated, when all data is 
taken together there is no significant trend in $N_{\rm  H}$ with redshift.
\begin{figure}
  \centering
  \resizebox{\hsize}{!}{ 
  \includegraphics[bbllx=67,bblly=320,bburx=548,bbury=743]{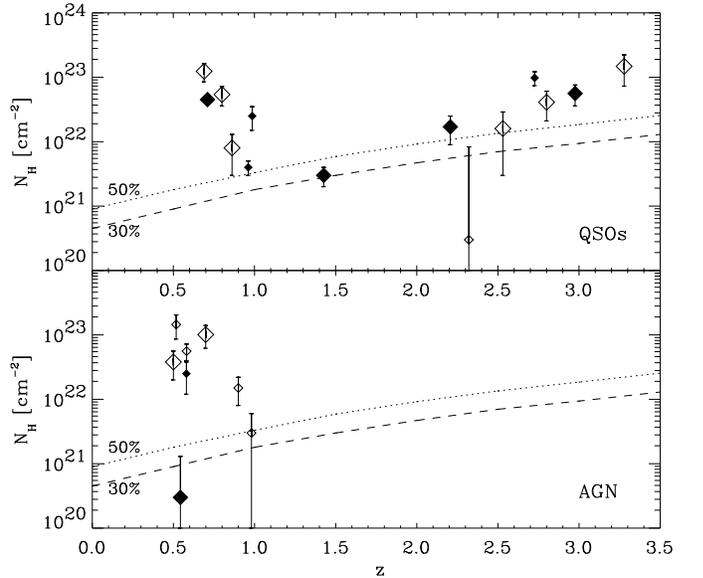}} 
      \caption{Intrinsic $N_{\rm H}$ vs. redshift diagram for
               type II QSOs (top panel) and type II AGN (lower panel). 
               Labels as in Fig.~\ref{fig:Nh_Lx}. 
               The dashed and dotted lines represent the 
               $N_{\rm H}$ values which correspond to a 30\% and 50\%
               decrease in the 0.5\,keV flux at the given redshift, respectively.}
         \label{fig:Nh_z}
\end{figure}
Considering the typical flux and column
densities found for our sources we expected all $z>1$ type II objects
to be classified as QSOs. At lower redshifts $\sim$40\% of the type II QSO 
candidates are actually type II QSOs. 

Statistical fluctuations in the X-ray spectrum can lead to high values 
of spuriously measured $N_{\rm H}$ values at high redshifts 
(e.g. \citealt{akylas}). However, Fig.~\ref{fig:diagnostic} show that 
objects with intrinsic $N_{\rm  H}$ of several 
$10^{22}-10^{23}$\,cm$^{-2}$ are not significantly 
influenced by systematic trends in redshift. The fit method is also able 
to pick up much higher absorptions than found in our sources.
The scatter in the mentioned   $N_{\rm  H}$ range at $z>2.5$ is consistent 
with an intrinsic absorption of $N_{\rm  H}=10^{23}$\,cm$^{-2}$ for all sources.

As a further test of the impact of statistical fluctuations, we plot in 
Fig.~\ref{fig:Nh_z} the amount of $N_{\rm  H}$ that is needed to reduce 
the 0.5\,keV flux by 30\% and 50\%. The 30\%-line agrees well with the 
90\% contours of unabsorbed sources in Fig.~\ref{fig:diagnostic}. 
Consequently, even for the sources at $z>2.5$ the fitted $N_{\rm  H}$ values
are unlikely to be caused by statistical fluctuations.

The previous conclusions depend on not having misinterpreted Compton-thick absorbed
objects ($N_{\rm  H} > 1.5 \times 10^{24}$\,cm$^{-2}$) as moderately absorbed objects.
Increasing column densities cause a hardening of the 
X-ray spectrum. This trend stops when the material becomes optically thick to
electron scattering.
The X-ray spectrum of a Compton-thick absorbed source is completely
dominated by the reflection component. Their X-ray spectra show soft X-ray
radiation but with a much lower photon index ($\Gamma \sim -0.4-1.3$, 
\citealt{maiolino}; \citealt{bassani}; \citealt{risaliti}; 
\citealt{iwasawa}; \citealt{comastri}).
We simulated an X-ray spectrum with $\Gamma=1.0$ 
as observed in the 3-12 keV band for NGC1068. When we fit the spectrum 
with a fixed $\Gamma=2.0$, we recover values for absorption of 
$N_{\rm H}=0.7\times 10^{22}$\,cm$^{-2}$, 
$N_{\rm H}=1.9\times 10^{22}$\,cm$^{-2}$ and 
$N_{\rm H}=3.8\times 10^{22}$\,cm$^{-2}$ for redshifts of $z=1,2,3$ respectively.
Hence, moderately absorbed
sources are found if intrinsically Compton-thick absorbed sources are studied.   
However, we are able to exclude a general misinterpretation of the spectra 
based on the following arguments.

First we tested if the X-ray spectra are better modelled with a purely 
reflection-dominated model ({\tt pexrav}). We used solar abundances and 
a cutoff energy of 100 keV. The ratio between the reflected and direct 
component (fit parameter: scaling factor) was set to 100 in order to obtain
pure reflection. The cosine of the inclination was left as a free parameter.
Again, we fitted the photon index as a 
free parameter only in the case the X-ray source had more than 100 PN net counts.  

Only the data of X-ray source X01135\_126 and X03246\_092 are slightly 
better represented by the reflection model in comparison to a single power 
law with intrinsic absorption. Further four X-ray sources (Marano 51A, 63A, 66A, 610A)
can be equally well modelled with a reflection model or a power law with 
intrinsic absorption (see Table~\ref{table2}). 
However, most of these objects have PN net counts of $<$50. 
Only X-ray source X01135\_126 has $N_{\rm PN-Counts} > 100$.
Low count spectra have the obvious problem that their X-ray data can be 
well fitted by different models. Nevertheless, the majority of our 
objects is only adequately fitted by a model that includes intrinsic
absorption.

Secondly the reflected component of Compton-thick absorbed sources is 
$\sim$50-150 times weaker than the de-absorbed intrinsic X-ray luminosity
(\citealt{brandt}). 
Hence they are 
usually found at low values of $f_{\rm X}/f_{\rm [OIII]}$ (\citealt{bassani}).
For the majority of our objects the [\ion{O}{iii}] line is redshifted out of
our spectral range. As a second best estimator we adopt 
$f_{\rm X}/f_{\rm OPT}$ rather than $f_{\rm X}/f_{\rm [OIII]}$.
Figure~\ref{fig:R_fx} shows that almost all
objects have high $f_{\rm X}/f_{\rm OPT}$ which is 
inconsistent with Compton-thick absorbed objects.
Only X00851\_154 has a rather low $f_{\rm X}/f_{\rm OPT}$ ratio, but
the best X-ray spectral fit is a moderately absorbed power law with $\Gamma=1.6$.
Under the assumption of Compton-thick
absorption, all studied objects would be classified as QSOs 
and our sample would contain the most X-ray luminous QSOs ever studied 
($L_{\rm{X_{INT}}}\sim 10^{47}\,{\rm erg/s}$).

\begin{figure}
  \centering
  \resizebox{\hsize}{!}{ 
  \includegraphics[bbllx=79,bblly=320,bburx=540,bbury=725]{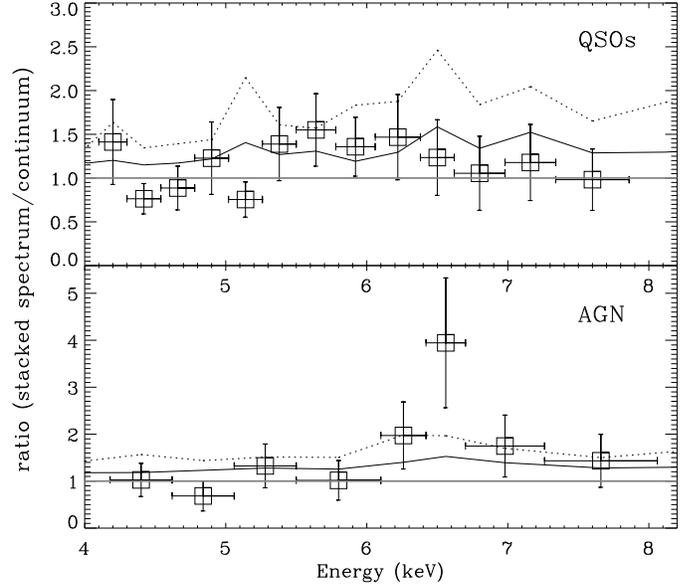}} 
   \caption{Stacked X-ray spectrum of all 14 type II QSOs (upper panel) 
            and 8 type II AGN (lower panel). The QSOs stack contains 
            $\sim$1600 counts in the shown 4-8.2\,keV energy range, the AGN 
            stack 300 counts. The 
            squares represent the ratio between the flux of the stacked
            spectrum and the simulated continuum spectrum. The solid line 
            shows the 1$\sigma$ detection error while the dotted line 
            represents the 2$\sigma$ detection error. The solid grey line at 
            a ratio of one illustrates the continuum flux.  }
          \label{fig:stacked}
\end{figure}

\subsection{Stacking of X-ray spectra\label{sec:stacking}}

Another indication for Compton-thick absorption is the detection of the iron line 
which is normally outshone by the direct component. 
Due to the reflection of the X-ray radiation from cold material in the torus,
an iron K$\alpha$ fluorescence line with an equivalent width $EW\gtrsim 1$\,keV 
is expected (\citealt{turner}, \citealt{risaliti2002}). 

The low count numbers in our X-ray spectra do not allow the study of
individual spectral features such as an iron line. Based on the assumption 
that the intrinsic X-ray properties in the sample are similar, 
a stack of the individual spectra can reveal the presence of an iron line. 
Since the number of individual X-ray spectra increases the SNR in the stacked 
spectrum, we decided to include all 14 type II QSOs regardless of the optical 
flag. The assumption is justified because the secure and tentative sample do 
not show differences in $f_{\rm X}/f_{\rm OPT}$ (Fig.~\ref{fig:R_fx})
and absorption values (Fig.~\ref{fig:Nh_hist}).

\cite{corral} explain the stacking process in detail. 
Here we outline the essential components of the process. Using the best-fit 
parameters, unfolded spectra ({\tt eufspec} in {\tt XSPEC}) from the ungrouped
observed X-ray spectra (MOS and PN data) are extracted. The unfolded spectra are corrected for
Galactic foreground absorption, shifted to the rest-frame and normalised to 
the same rest-frame flux in the 2-5 keV band. The flux is rebinned into 
a common energy grid for all spectra. The stacked spectrum is binned to 
a final energy grid that contains at least 100 counts per energy bin. The 
final errors of the 
stacked spectrum are based on Gaussian propagation of the errors in the
individual spectra. To distinguish between real spectral features and
artifacts from the averaging process the underlying continuum has to be 
modelled. Each source was simulated 100 times using the same model (absorption
plus powerlaw) with the same 2-8\,keV flux as observed in the real spectra. 
The individual simulated X-ray continuum spectra from all different X-ray 
sources are stacked exactly as the observed X-ray spectra which are used to
determine the 1$\sigma$ and 2$\sigma$ contours (68\% and 95\% of all simulated
continuum spectra).

The 0.5-10\,keV final stacked spectrum of the QSOs contains $\sim$4300
counts. Together with the 1$\sigma$ and 2$\sigma$ contours the
ratio of the observed spectrum to the average simulated continuum 
is shown in Fig.~\ref{fig:stacked}, upper panel. 
The stacked type II QSO spectrum shows no residuals around 6-7\,keV 
which would correspond to an iron K$\alpha$ emission line.
The SNR in the stacked spectrum is sufficient to detect an 
iron line with $EW\gtrsim 1$\,keV. The stacked spectrum does not 
provide any evidence for Compton-thick absorbed sources based on a line emission 
and supports the assumption that our type II QSOs are actually moderately 
absorbed sources.

Interestingly, the stacked spectrum of our 8 type II AGN does show a 
very prominent emission line (Fig.~\ref{fig:stacked}, lower panel), although in 
none of the individual AGN X-ray spectra a significant excess 
at 6-7\,keV is recognised (see the X-ray spectra in the 
Appendix~\ref{appendix1}). Since only a few objects are included in the 
AGN stack we used energy bins containing 30 counts instead of 100 counts. 
We fitted the positive residuals in the stacked AGN spectrum with 
a Gaussian line profile. A line is detected with a significance 
above 2$\sigma$ at a line energy of $E=6.6_{-0.08}^{+0.07}$\,keV 
with a $\sigma=230_{-40}^{+100}$\,eV. The equivalent width of the fit is
$EW=2.0_{-0.7}^{+0.6}$\,keV.

In principle there are two possible explanations for the detection 
of a broad iron fluorescence line. 
\begin{itemize}
 \item{a reflection line from the accretion disk, which
  can be significantly broadened due to relativistic effects}
 \item{a blend of narrow lines that is observed as a single unresolved 
       broad line due to the low signal-to-noise ratio of our observations}
\end{itemize}
The typical equivalent widths of the disk iron lines
are usually significantly lower than the $EW$ of the iron line found in our 
AGN stack. The energy and width of the line is more consistent with a blend of
narrow lines. Depending on the properties of the reflecting material 
different ionisation stages of iron are excited as observed in the 
the Compton-thick absorbed object NGC 1068 (\citealt{ogle}). 
If the signal-to-noise ratio is lowered the three resolved lines will be observed as an
unresolved broad ($EW\gtrsim 1$\,keV) line with a peak above 6.4\,keV. 
In this case a significant number of objects that we classified as AGN
could be Compton-thick absorbed. Interestingly, except for one object, all 
X-ray sources, that can be better or equally well fitted by a purely 
reflection-dominated model in comparison to the power law plus intrinsic absorption
model, are AGN.

\section{Conclusions\label{conclusions}}
We selected 51 spectroscopically classified type II objects from the
{\it XMM-Newton} Marano field survey, XWAS, and AXIS. We 
re-investigate the optical spectra for narrow and common AGN high 
excitation lines, as well as verifying the X-ray counterpart determination. 
22 sources meet our selection criteria (13 secure identification, 
9 tentative). All sources have $z\ge 0.5$ and observed 0.5-10\,keV 
X-ray luminosities (not corrected for intrinsic absorption) of 
$L_{\rm X_{\rm OBS}}\sim 10^{43-45}$\,erg/s. The sample is not flux limited. 
 
The selection of both optical, narrow high
excitation emission lines and intrinsic X-ray luminosities 
$L_{\rm X_{\rm INT}}\geq 10^{44}$\,erg/s yielded 14 type II QSOs.

Since the distribution of net PN counts peaks at $\sim$40 the \mbox{X-ray} spectral
analysis has to account for this very low numbers of counts. We
extensively simulated and studied such low count number X-ray spectra. 
A binning with at least one count per bin combined with the Cash-statistic recovered the
input values best. We proved that the method is able to find 
absorption up to $N_{\rm H}=10^{24}$\,cm$^{-2}$.

However, we discover only moderately absorbed type II QSOs. One QSO
is consistent with being unabsorbed but it belongs to the tentative sample. 
Compton-thick absorbed sources may be detected as moderately
absorbed but we can exclude this scenario for the majority of our 
sources. The X-ray data are not well fitted by reflection models. 
The $f_{\rm X}/f_{\rm OPT}$ values of the objects and the non-detection of 
a broad iron line in the stacked type II QSOs spectrum give evidence that 
we have not misclassified Compton-thick absorbed sources. In contrary to the QSO stack, 
the stack of 8 type II AGN revealed a very prominent iron line with an 
$EW\sim 2$\,keV. However, the shape of the single X-ray spectra and the
modelling of the X-ray data showed that the majority of our AGN are most
probably moderately absorbed X-ray sources.

The column density distribution found by us agrees well with those 
in deep {\it Chandra} and {\it XMM-Newton} surveys (\citealt{mainieri};
\citealt{szokoly}; \citealt{mateos}; \citealt{ptak}). These authors also reported a few 
cases of unabsorbed type II AGNs,
as well as evidence for some heavily absorbed sources ($N_{\rm H}\sim
10^{24}$\,cm$^{-2}$). \cite{lafranca} studied column density trends
with X-ray luminosity in different redshift bins. They
classified type II AGN as all sources that do not show any emission lines 
with FWHM $>$ 2000\,km/s. Independent of X-ray luminosity and redshift bins, 
they discovered type II AGN with an average absorption of $N_{\rm H}\sim
10^{23}$\,cm$^{-2}$, slightly above the column density peak of our
survey ($N_{\rm H}= 4 \times 10^{22}$\,cm$^{-2}$). The 
present study of strictly classified type II AGN/QSOs, based on the 
optical spectra, verified that there is no obvious trend of absorbing column
density with redshift or X-ray luminosity.

Our results apparently contradict studies of the local universe. \cite{bassani} and 
\cite{risaliti} found 75\% of their sources with high absorption 
($N_{\rm  H} > 10^{23}$\,cm$^{-2}$) and 25-45\% with 
$N_{\rm  H} > 10^{24}$\,cm$^{-2}$. Either the column density properties change
dramatically from the local to distant universe or the majority of the 
heavily absorbed distant sources are missed in our surveys. Indeed  
even the most luminous Compton-thick absorbed sources in the local universe, 
e.g. NGC 6240, could not have been detected by our survey 
if they were at $z \gtrsim 0.4$.

In summary, the column densities in our survey show no trend in 
X-ray luminosity and no clear trend in redshift. 
If we compare our results with samples of AGN in the local universe,
our sample does not contain a significant fraction of heavily absorbed sources 
($N_{\rm  H} > 10^{23}$\,cm$^{-2}$).
Our survey of objects with $z\ge 0.5$ is limited to observed X-ray luminosities 
in excess of
$L_{\rm{X_{OBS}}}> 10^{43}$\,erg/s. We can only expect to find QSOs
intrinsically more luminous than $L_{\rm{X_{INT}}}> 10^{45}$\,erg/s, 
if they are Compton-thick absorbed. \cite{norman} claim that for 
the most distant type II QSO ever detected ($z=3.7$), there is strong evidence for 
heavy or even Compton-thick absorption. However, our survey
rules out large numbers of Compton-thick absorbed sources with X-ray 
luminosities of $L_{\rm{X_{INT}}}> 10^{45}$\,erg/s. Hence, potential Compton-thick 
absorbed objects at high redshifts are likely to have similar X-ray
luminosities to Compton-thick absorbed objects in the local universe. 
In order to find a supposed, rare population of very luminous, Compton-thick 
absorbed QSOs a larger survey area is needed. The 2XMM catalogue 
 (\citealt{watson}) with a survey area of $\sim$360\,deg$^2$ could provide 
a valuable source to reveal such a population.

\begin{acknowledgements} 
Mirko Krumpe is supported by the Deutsches
Zentrum f\"ur Luft- und Raumfahrt (DLR) GmbH  
under contract No. FKZ 50 OR 0404.
Georg Lamer acknowledges support by the Deutsches Zentrum f\"ur Luft- und  
Raumfahrt (DLR) GmbH under contract no.~FKZ 50 OX 0201. Amalia Corral,
Francisco J. Carrera, and Xavier Barcons acknowledge financial support by the
Spanish Ministry of Education and Science, through projects
ESP2006-13608-C02-01. MP, SM, JT and MW thank STFC for financial support.
\end{acknowledgements}

\end{twocolumn}
\Online
\onecolumn
\appendix 

\section{Optical \& X-ray spectra, confidence contours\label{appendix1}}
In this section we show the optical and X-ray spectra, as well as the 
confidence contours of X-ray spectral fits for the parameters $N_{\rm H}$ and
$\Gamma$.\\

{\em Optical spectra:} Optical atmospheric absorption corrected, wavelength and 
flux calibrated spectrum for the optical X-ray counterpart is shown. All spectra 
are in
flux units of $10^{-18}$\,erg\,cm$^{-2}$\,s$^{-1}$\,\AA$^{-1}$. The black solid 
line represents the spectrum, the green solid line the error spectrum (not
available in all spectra). Red markers indicate possible emission lines, 
blue markers absorption lines. There are exceptions in a few spectra.
Spectral features at 5580\,\AA\ are spurious due to incomplete subtraction 
of a night sky line.\\

{\em X-ray spectra:} 

The X-ray spectra show the PN (black) and combined MOS
(red) data, as
well as the best fit model (foreground Galactic absorption, power law with
intrinsic absorption at the object's redshift). In the case of Marano 51A and
Marano 610A the shown best fit model consists of foreground Galactic
absorption, an unabsorbed power law, and an power law with intrinsic
absorption at the object's redshift to account for the soft excess in the
X-ray data.

For objects that have less than 100 PN source
counts in the 0.2-8\,keV band we fixed the photon index
($\Gamma=2$). Otherwise, free fits in $N_{\rm H}$ and $\Gamma$ are shown. 
The fit parameters are given in Table~\ref{table2}. For illustration
purposes the X-ray data have been rebinned to different signal-to-noise ratio, 
after grouping to a minimum of one count per bin (min 1).\\

{\em Confidence contours:} Confidence contours of the absorbing column density in units
of $N_{\rm H}= 10^{22}$\,cm$^{-2}$ vs. photon index $\Gamma$ of the X-ray
spectral fits. Confidence contours are plotted for 68\% (black), 90\% (red) and 99\%
(green). The contours are based on free $N_{\rm H}$ and $\Gamma$ X-ray spectral 
fits for all objects independent of PN net counts.\\

{\em Comments:} Below every set of optical \& X-ray spectra and X-ray contour a 
comment for the objects shown is given. This comment includes the classification 
number of the optical counterpart and the corresponding redshift. Furthermore,
a short fit description is given. For more details on the properties of the
sources and the X-ray spectral fits see Tables~\ref{table1}, \ref{table2}.

\setlength{\unitlength}{1.0cm}
\hspace*{1cm}

\begin{picture}(10,10)
 \put(4.2,9.6){\epsfig{file=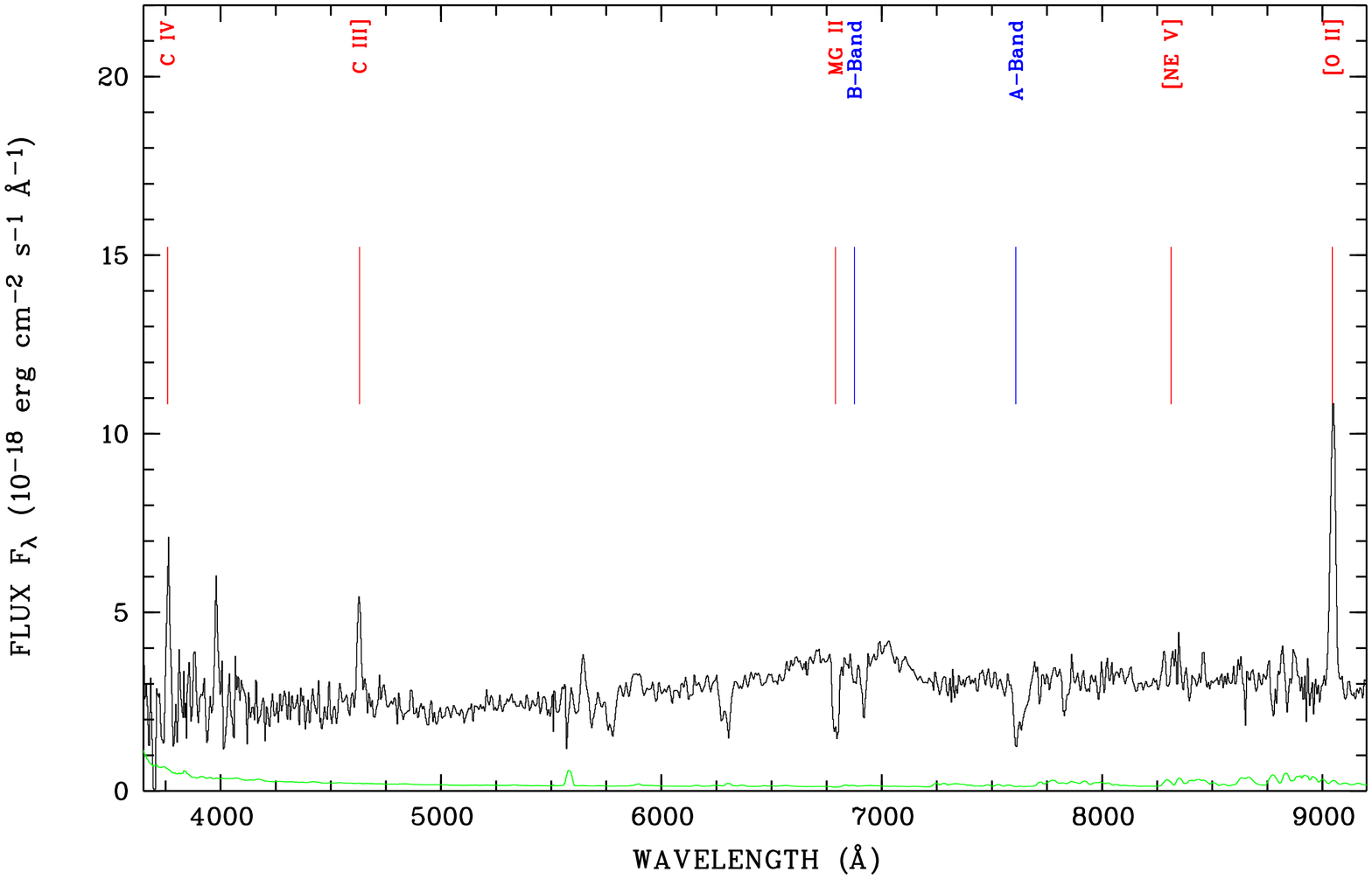,width=6.5cm,angle=270}}
 \put(0.2,3.4){\epsfig{file=spectra/Xray_spectrum_abs_9.ps,width=6cm,angle=270}}
 \put(9.2,3.4){\epsfig{file=spectra/Xray_contour_9.ps,width=6cm,angle=270}}
 \put(5.4,-3.0){object Marano 9A, z = 1.427, free X-ray spectrum fit}
\end{picture} 

\newpage
\begin{picture}(10,23)
 \put(4.2,23.0){\epsfig{file=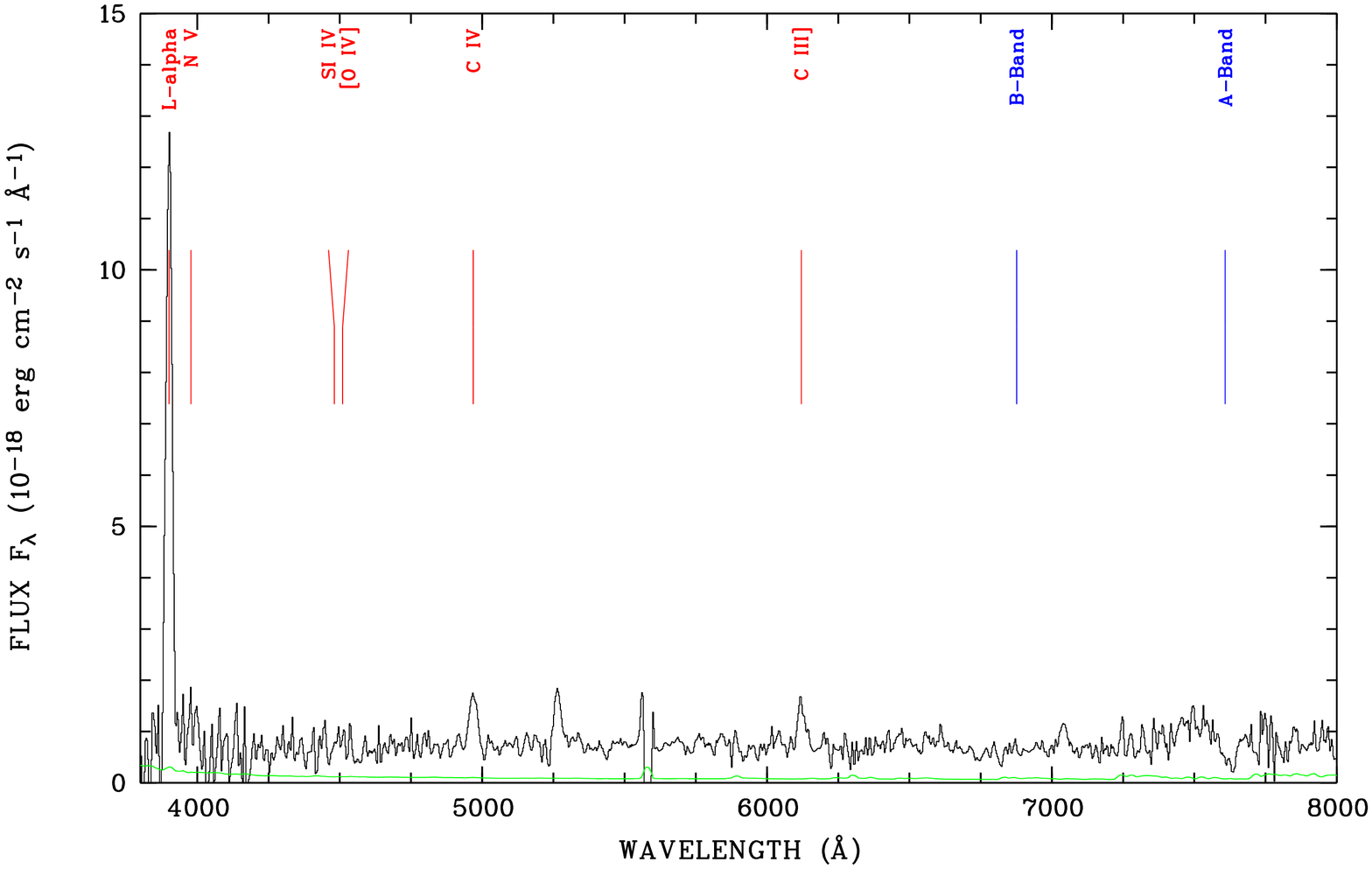,width=6.5cm,angle=270}}
 \put(0.2,16.8){\epsfig{file=spectra/Xray_spectrum_abs_20.ps,width=6cm,angle=270}}
 \put(9.2,16.8){\epsfig{file=spectra/Xray_contour_20.ps,width=6cm,angle=270}}
 \put(4.9,10.4){object Marano 20A,  z = 2.207, frozen $\Gamma=2.0$ X-ray spectrum fit}

 \put(4.2,10){\epsfig{file=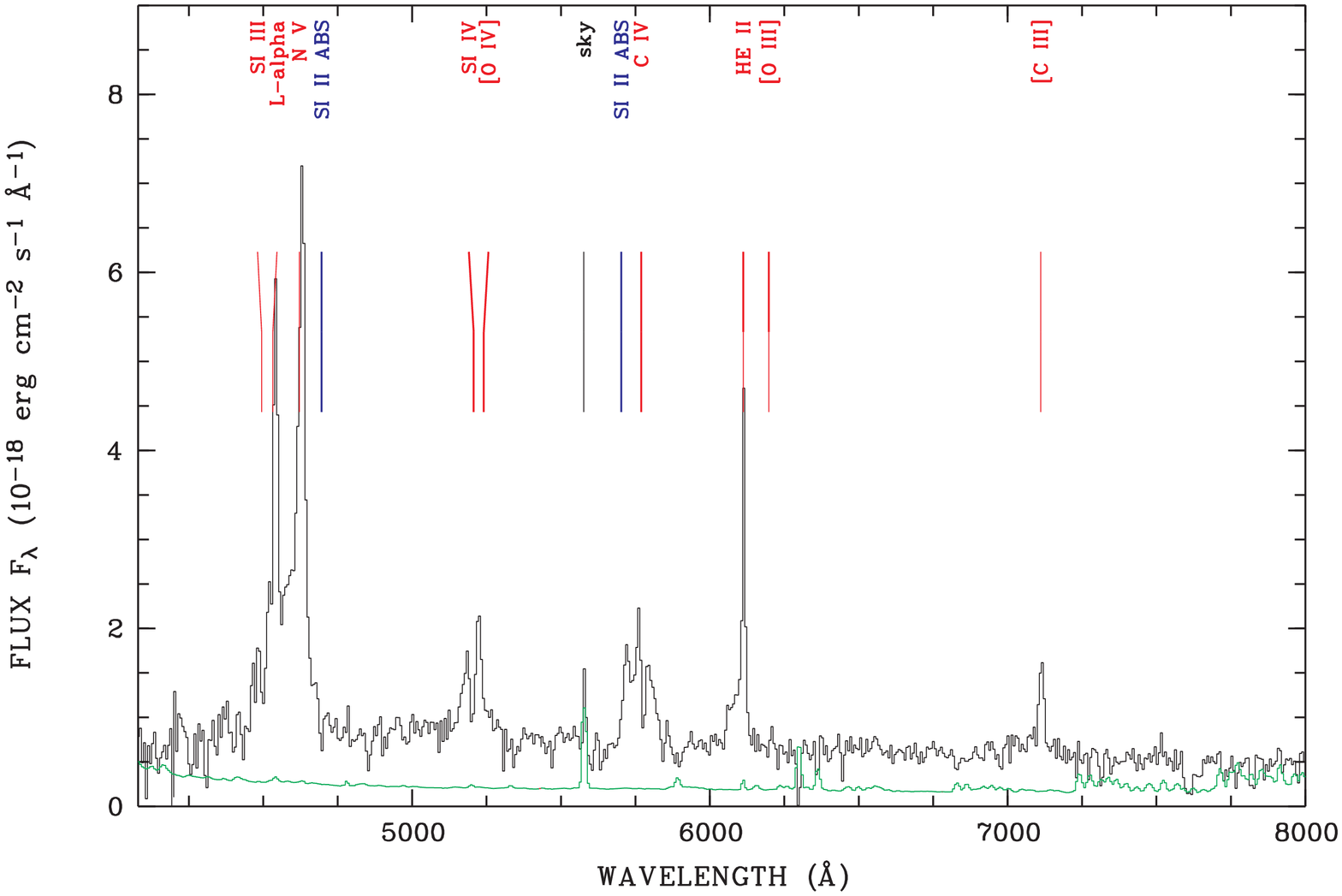,width=6.5cm,angle=270}}
 \put(0.2,3.8){\epsfig{file=spectra/Xray_spectrum_abs_32.ps,width=6cm,angle=270}}
 \put(9.2,3.8){\epsfig{file=spectra/Xray_contour_32.ps,width=6cm,angle=270}}
 \put(5.4,-2.6){object Marano 32A,  z = 2.727, free X-ray spectrum fit}
\end{picture} 

\newpage
\begin{picture}(10,23)
 \put(4.2,23.0){\epsfig{file=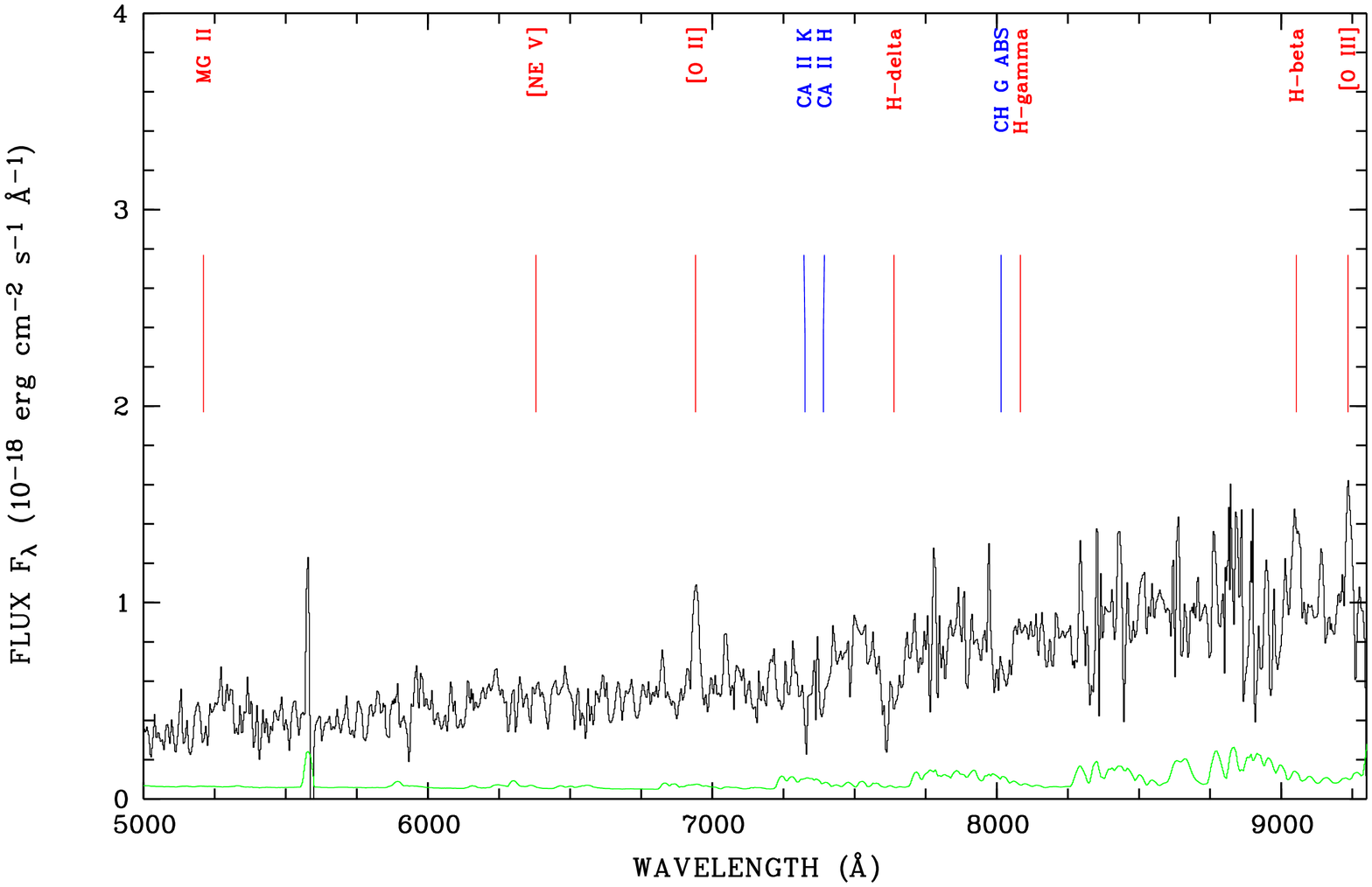,width=6.5cm,angle=270}}
 \put(0.2,16.8){\epsfig{file=spectra/Xray_spectrum_abs_39.ps,width=6cm,angle=270}}
 \put(9.2,16.8){\epsfig{file=spectra/Xray_contour_39.ps,width=6cm,angle=270}}
 \put(4.9,10.4){object Marano 39A,  z = 0.862, frozen $\Gamma=2.0$ X-ray spectrum fit}

 \put(4.2,10){\epsfig{file=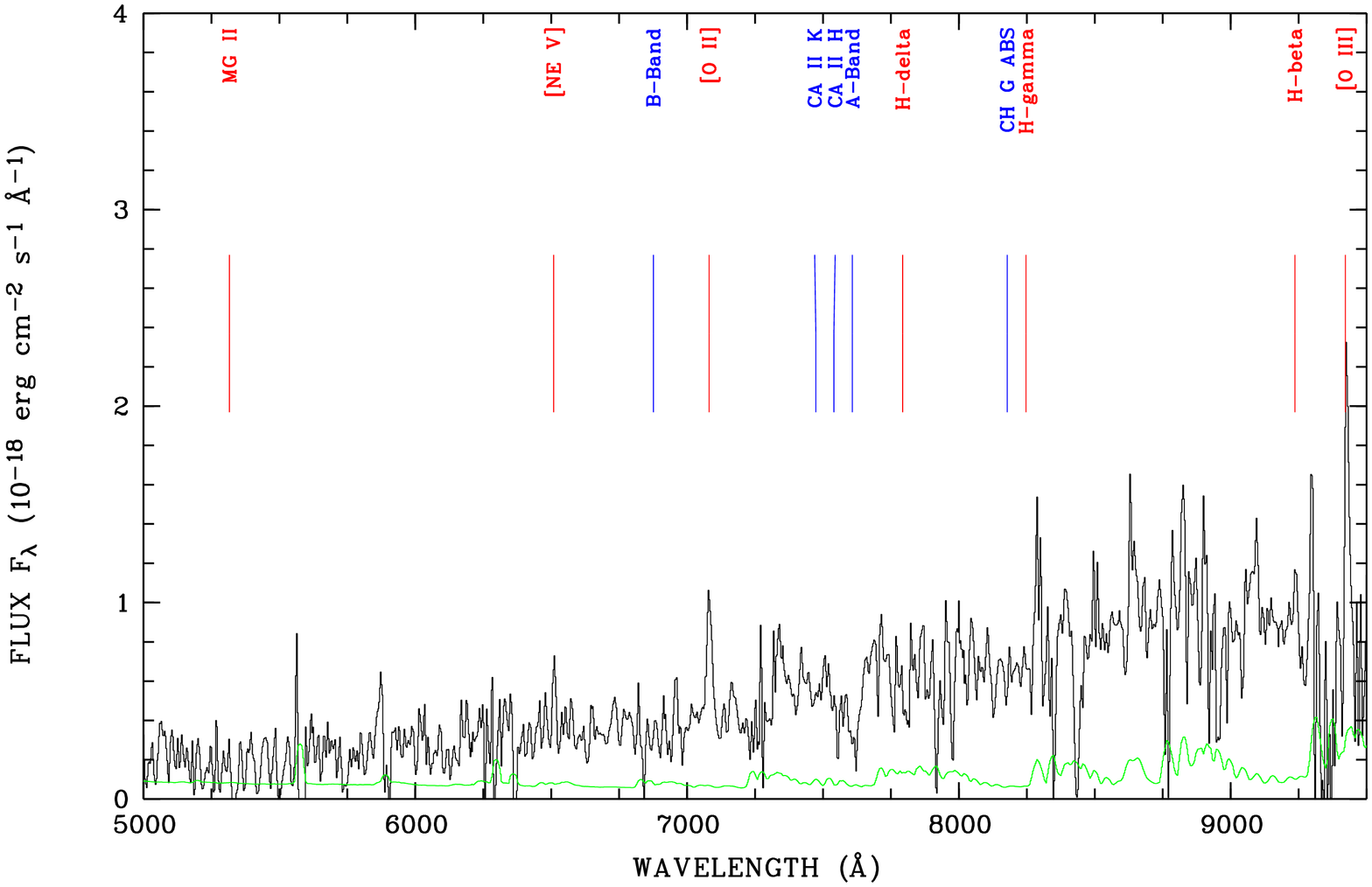,width=6.5cm,angle=270}}
 \put(0.2,3.8){\epsfig{file=spectra/Xray_spectrum_abs_47.ps,width=6cm,angle=270}}
 \put(9.2,3.8){\epsfig{file=spectra/Xray_contour_47.ps,width=6cm,angle=270}}
 \put(4.9,-2.6){object Marano 47A,  z = 0.900, frozen $\Gamma=2.0$ X-ray spectrum fit}
\end{picture} 

\newpage
\begin{picture}(10,23)
 \put(4.2,23.0){\epsfig{file=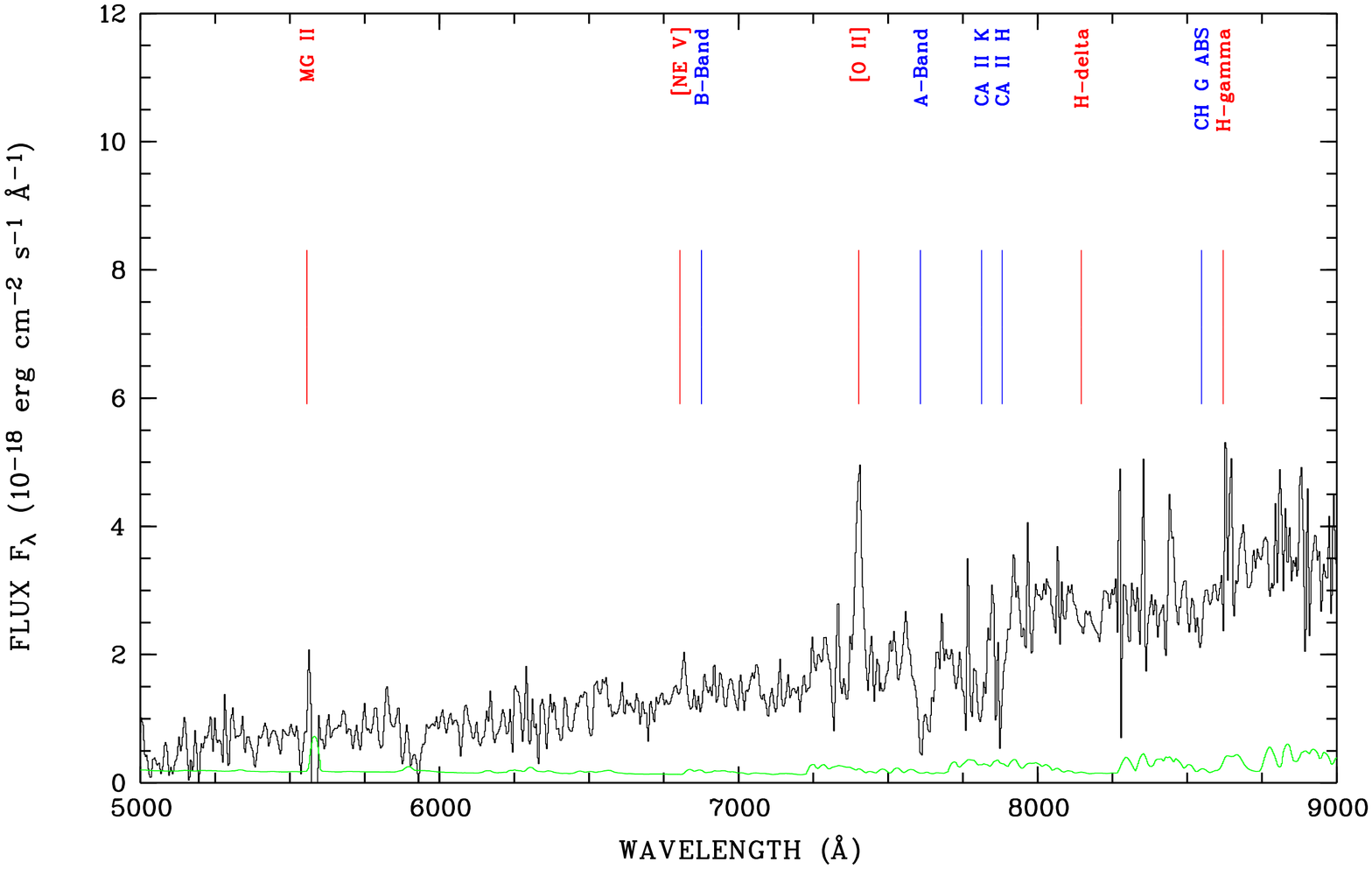,width=6.5cm,angle=270}}
 \put(0.2,16.8){\epsfig{file=spectra/Xray_spectrum_abs_50.ps,width=6cm,angle=270}}
 \put(9.2,16.8){\epsfig{file=spectra/Xray_contour_50.ps,width=6cm,angle=270}}
 \put(4.9,10.4){object Marano 50A,  z = 0.986, frozen $\Gamma=2.0$ X-ray spectrum fit}

 \put(4.2,10.0){\epsfig{file=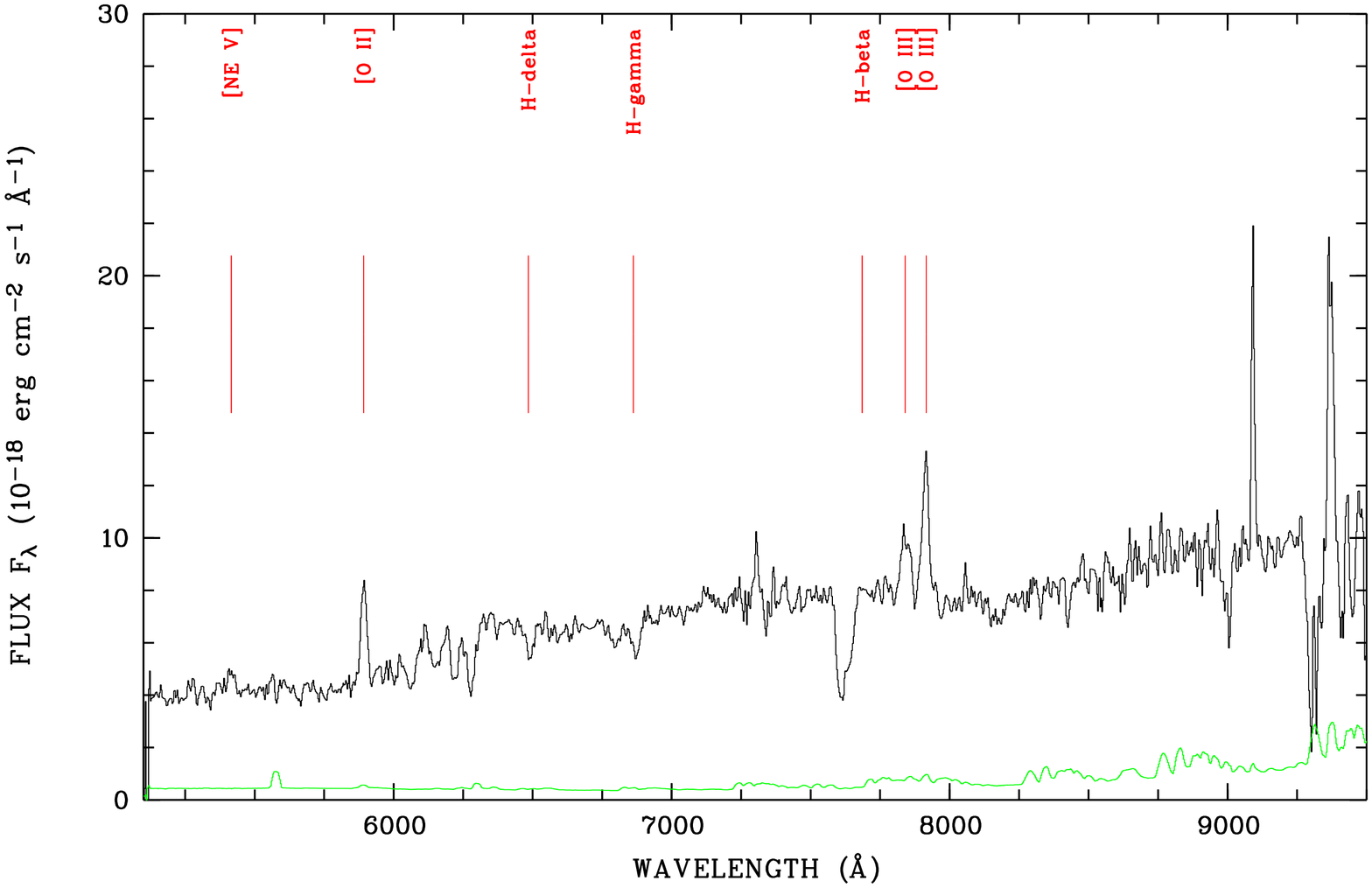,width=6.5cm,angle=270}}
 \put(0.2,3.8){\epsfig{file=spectra/Xray_spectrum_pow_zwabspow_51.ps,width=6cm,angle=270}}
 \put(9.2,3.8){\epsfig{file=spectra/Xray_contour_pow_zwabspow_51.ps,width=6cm,angle=270}}
 \put(3.9,-2.6){object Marano 51A,  z = 0.58, two power laws, frozen $\Gamma=2.0$ X-ray spectrum fit}
\end{picture} 

\newpage
\begin{picture}(10,23)
 \put(4.2,23){\epsfig{file=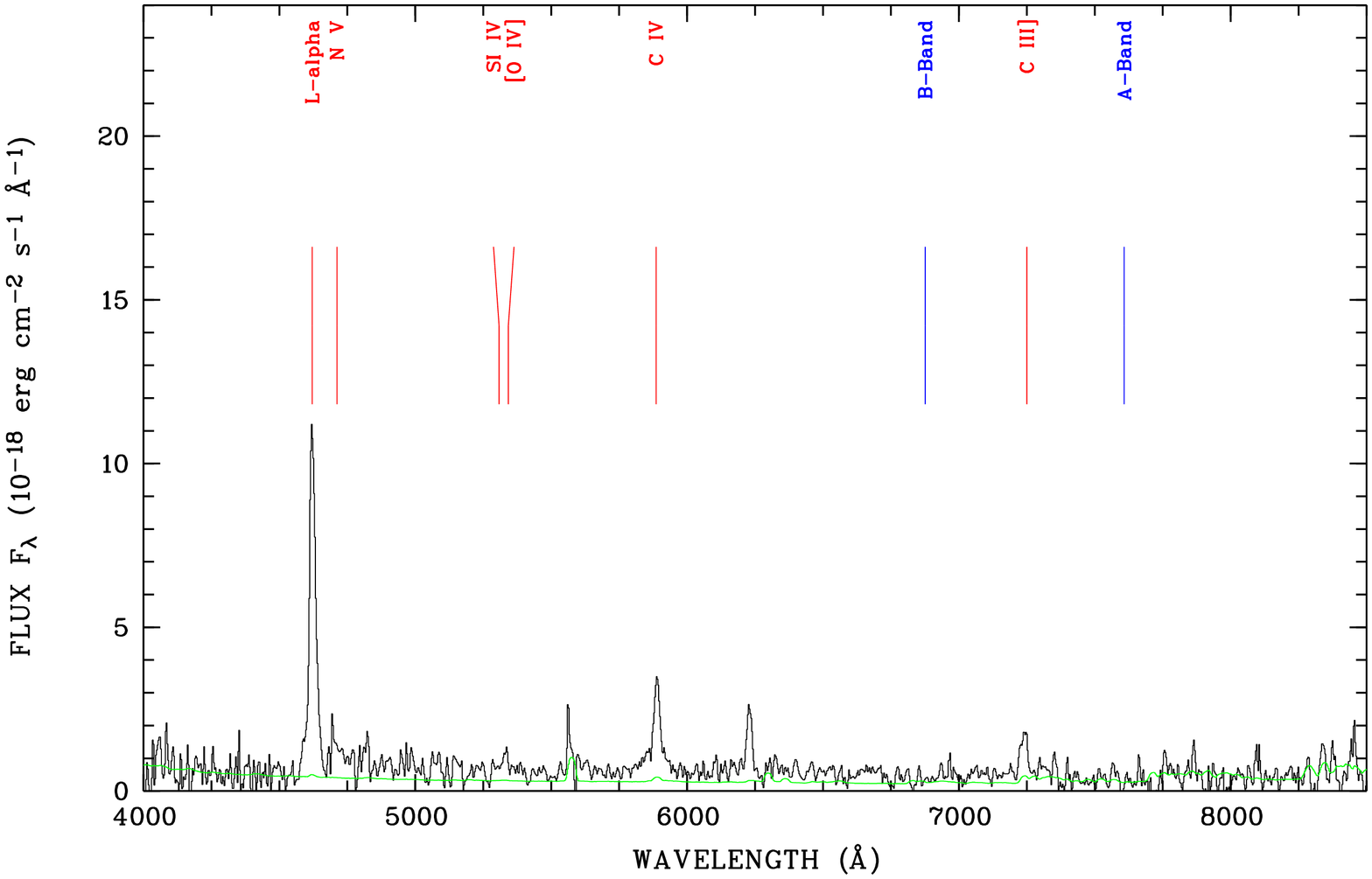,width=6.5cm,angle=270}}
 \put(0.2,16.8){\epsfig{file=spectra/Xray_spectrum_abs_63.ps,width=6cm,angle=270}}
 \put(9.2,16.8){\epsfig{file=spectra/Xray_contour_63.ps,width=6cm,angle=270}}
 \put(4.9,10.4){object Marano 63A,  z = 2.800, frozen $\Gamma=2.0$ X-ray spectrum fit}

 \put(4.2,10){\epsfig{file=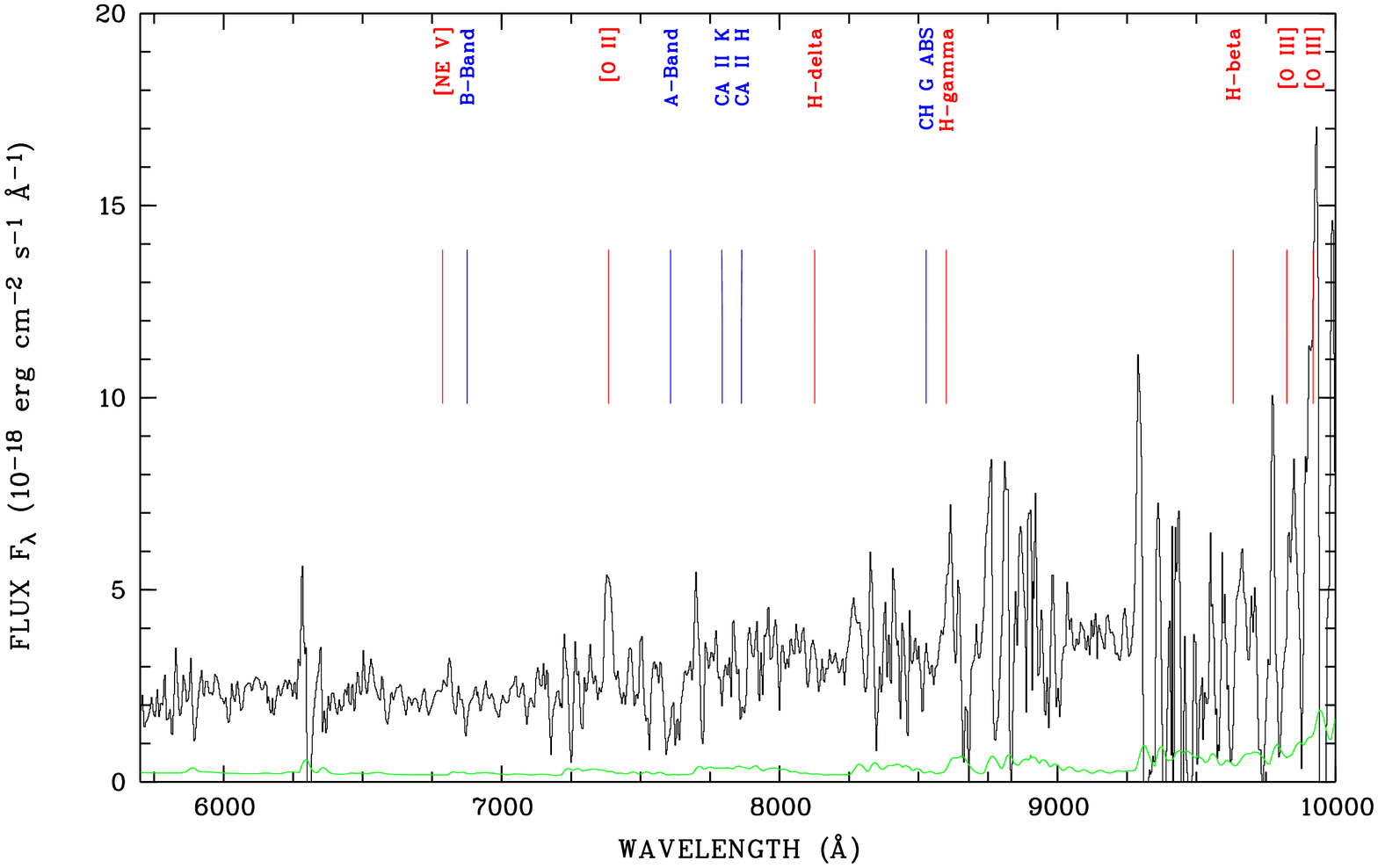,width=6.5cm,angle=270}}
 \put(0.2,3.8){\epsfig{file=spectra/Xray_spectrum_abs_66.ps,width=6cm,angle=270}}
 \put(9.2,3.8){\epsfig{file=spectra/Xray_contour_66.ps,width=6cm,angle=270}}
 \put(4.9,-2.6){object Marano 66A,  z = 0.981, frozen $\Gamma=2.0$ X-ray spectrum fit}
\end{picture} 

\newpage
\begin{picture}(10,23)
 \put(4.2,23){\epsfig{file=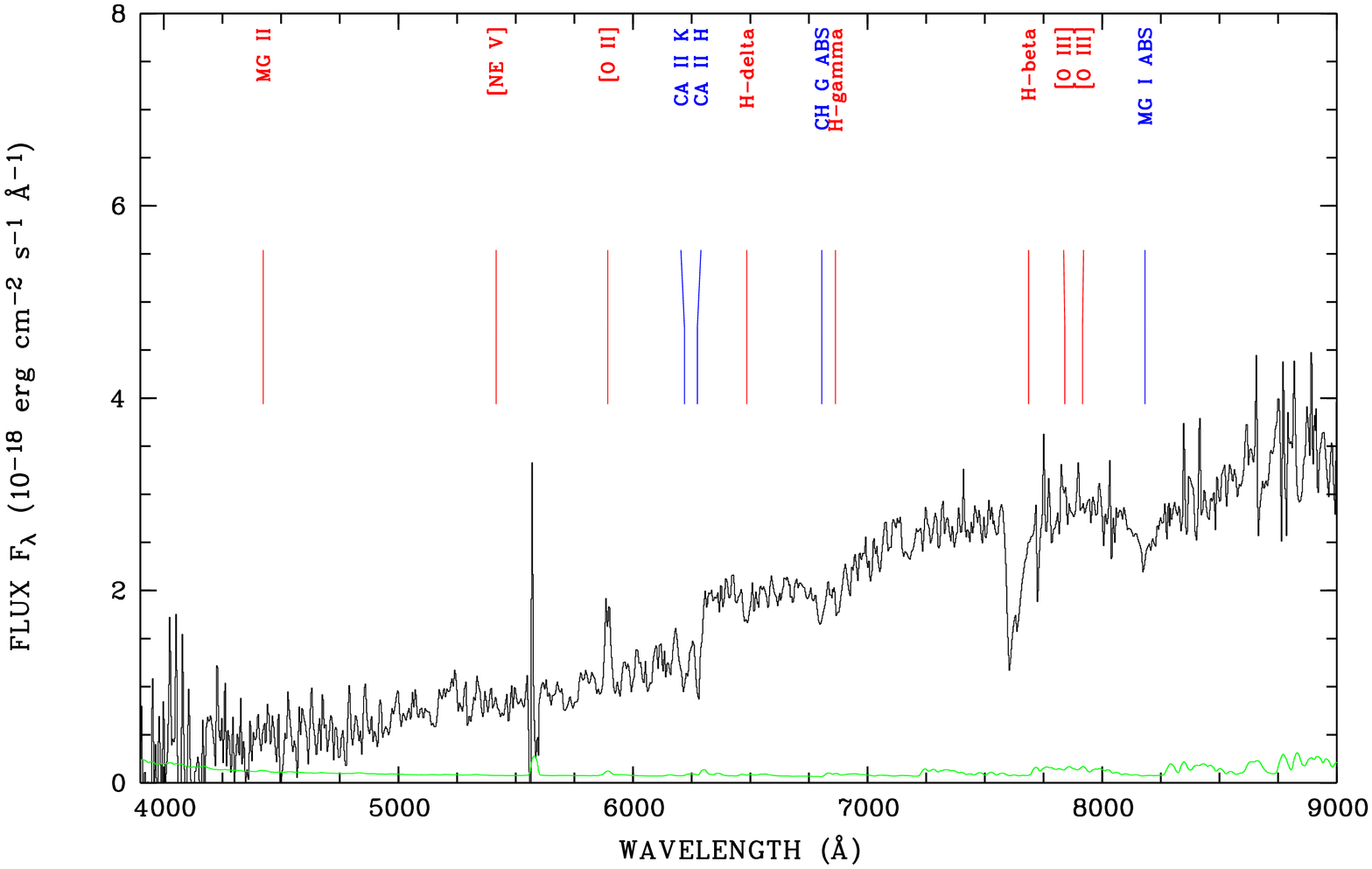,width=6.5cm,angle=270}}
 \put(0.2,16.8){\epsfig{file=spectra/Xray_spectrum_abs_116.ps,width=6cm,angle=270}}
 \put(9.2,16.8){\epsfig{file=spectra/Xray_contour_116.ps,width=6.7cm,angle=270}}
 \put(4.9,10.4){object Marano 116A,  z = 0.581, frozen $\Gamma=2.0$ X-ray spectrum fit}

 \put(4.2,10){\epsfig{file=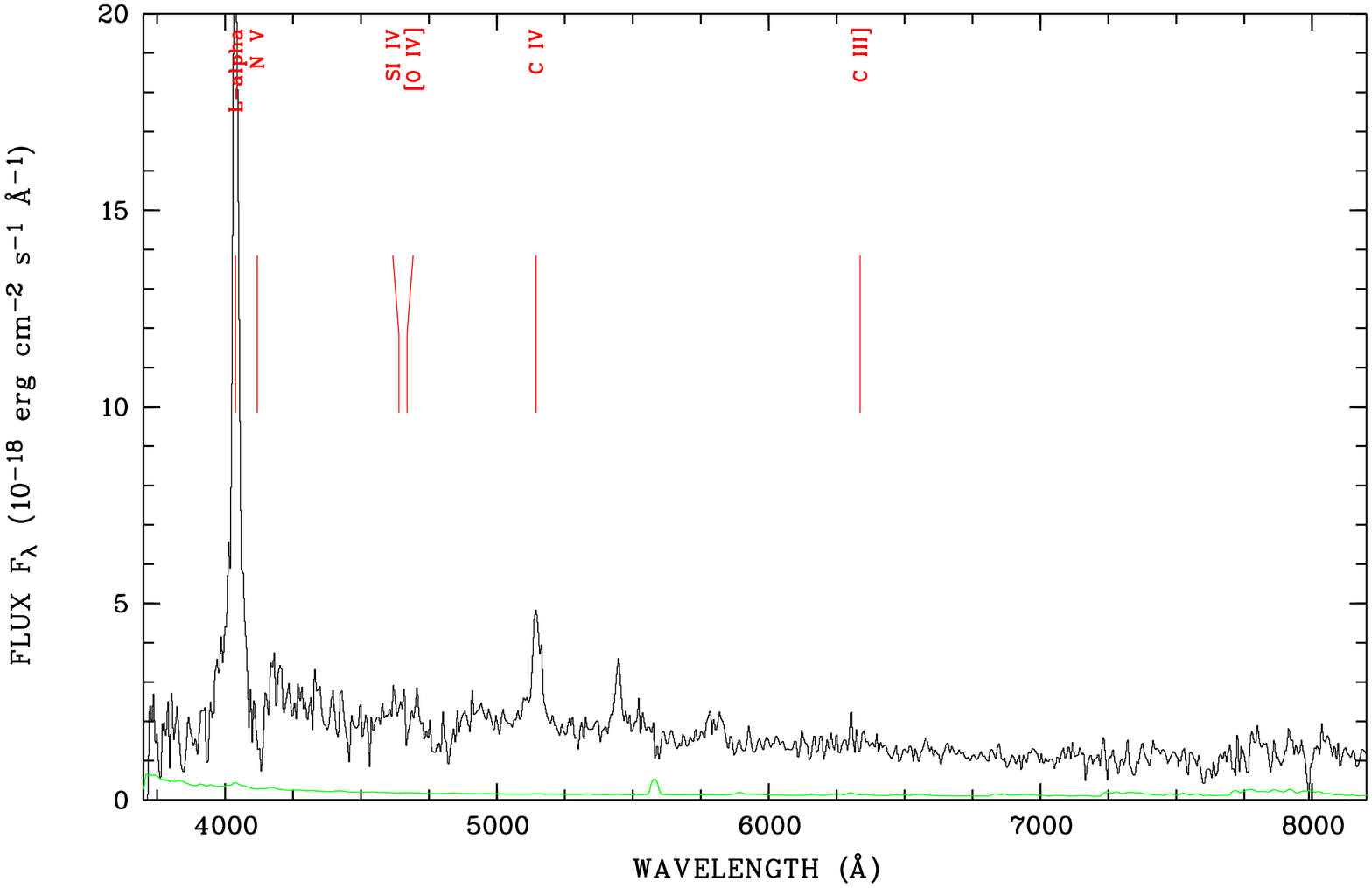,width=6.5cm,angle=270}}
 \put(0.2,3.8){\epsfig{file=spectra/Xray_spectrum_abs_133.ps,width=6cm,angle=270}}
 \put(9.2,3.8){\epsfig{file=spectra/Xray_contour_133.ps,width=6cm,angle=270}}
 \put(4.9,-2.6){object Marano 133A,  z = 2.321, frozen $\Gamma=2.0$ X-ray spectrum fit}
\end{picture} 

\newpage
\begin{picture}(10,23)
 \put(4.2,23){\epsfig{file=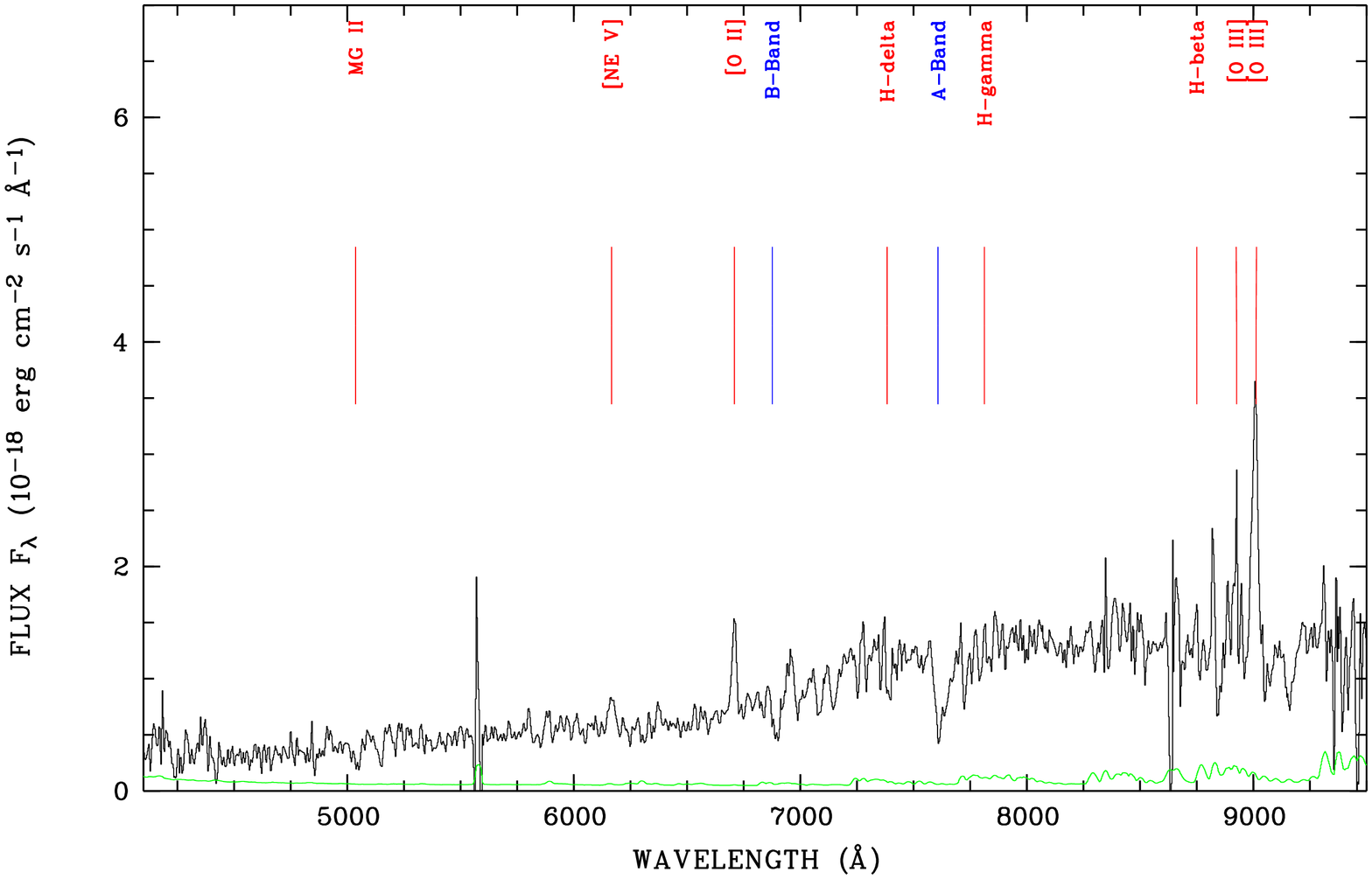,width=6.5cm,angle=270}}
 \put(0.2,16.8){\epsfig{file=spectra/Xray_spectrum_abs_171.ps,width=6cm,angle=270}}
 \put(9.2,16.8){\epsfig{file=spectra/Xray_contour_171.ps,width=6cm,angle=270}}
 \put(4.9,10.4){object Marano 171A,  z = 0.800, frozen $\Gamma=2.0$ X-ray spectrum fit}

 \put(4.2,10){\epsfig{file=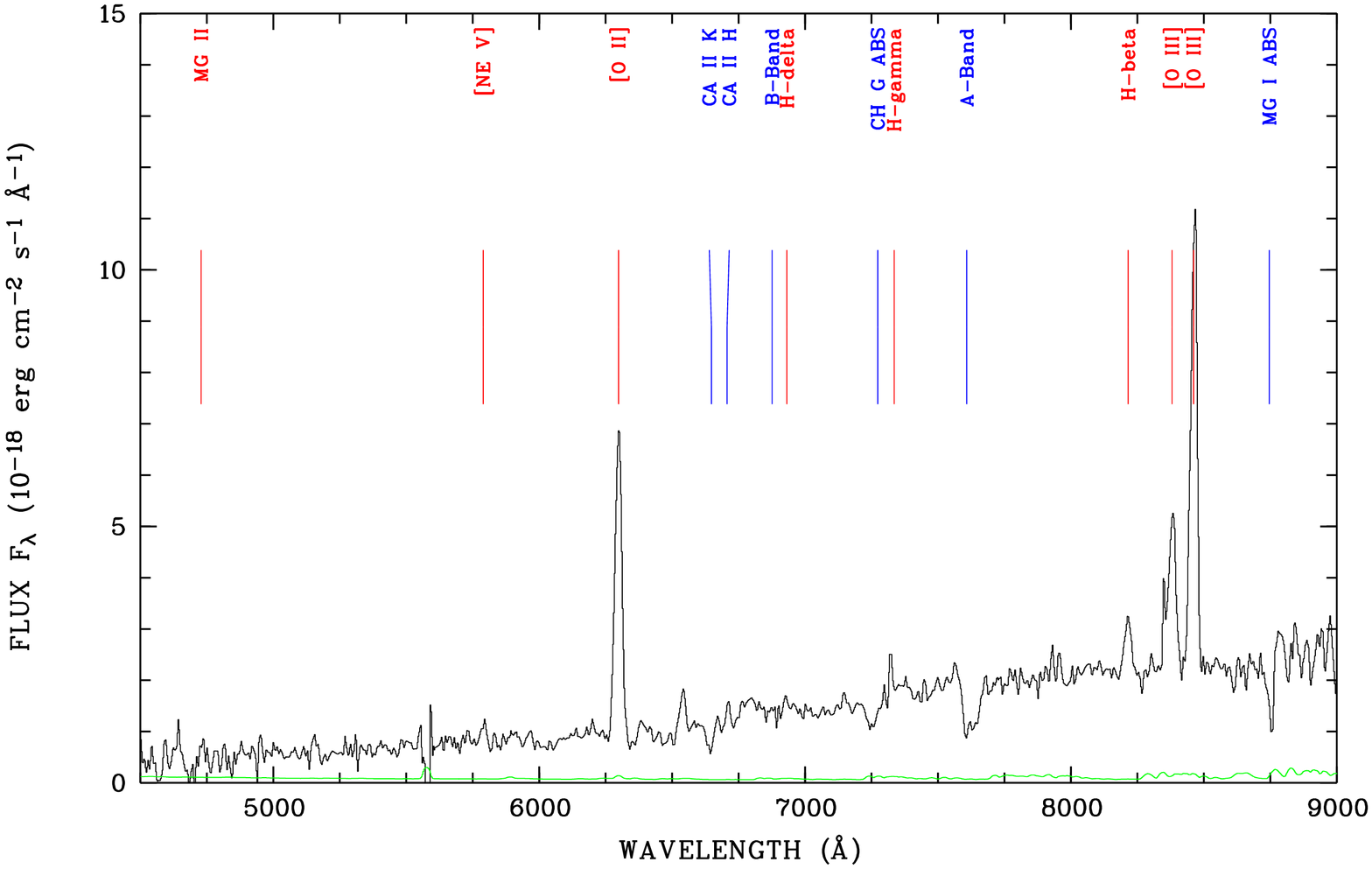,width=6.5cm,angle=270}}
 \put(0.2,3.8){\epsfig{file=spectra/Xray_spectrum_abs_224.ps,width=6cm,angle=270}}
 \put(9.2,3.8){\epsfig{file=spectra/Xray_contour_224.ps,width=6cm,angle=270}}
 \put(4.9,-2.6){object Marano 224B,  z = 0.690, frozen $\Gamma=2.0$ X-ray spectrum fit}
\end{picture} 

\newpage
\begin{picture}(10,23)
 \put(4.2,23){\epsfig{file=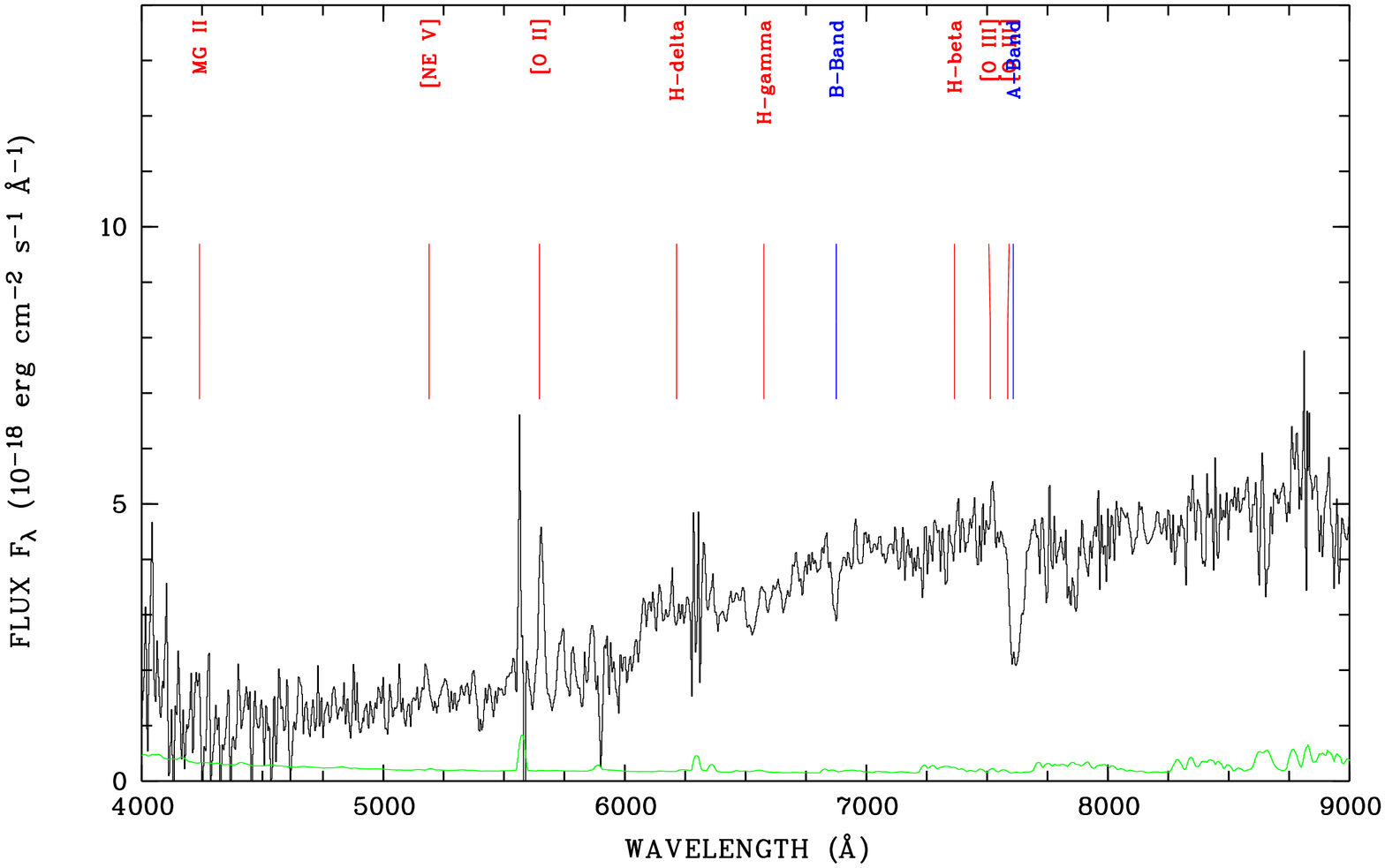,width=6.5cm,angle=270}}
 \put(0.2,16.8){\epsfig{file=spectra/Xray_spectrum_abs_253.ps,width=6cm,angle=270}}
 \put(9.2,16.8){\epsfig{file=spectra/Xray_contour_253.ps,width=6.4cm,angle=270}}
 \put(4.9,10.4){object Marano 253A,  z = 0.517, frozen $\Gamma=2.0$ X-ray spectrum fit}

 \put(4.2,10){\epsfig{file=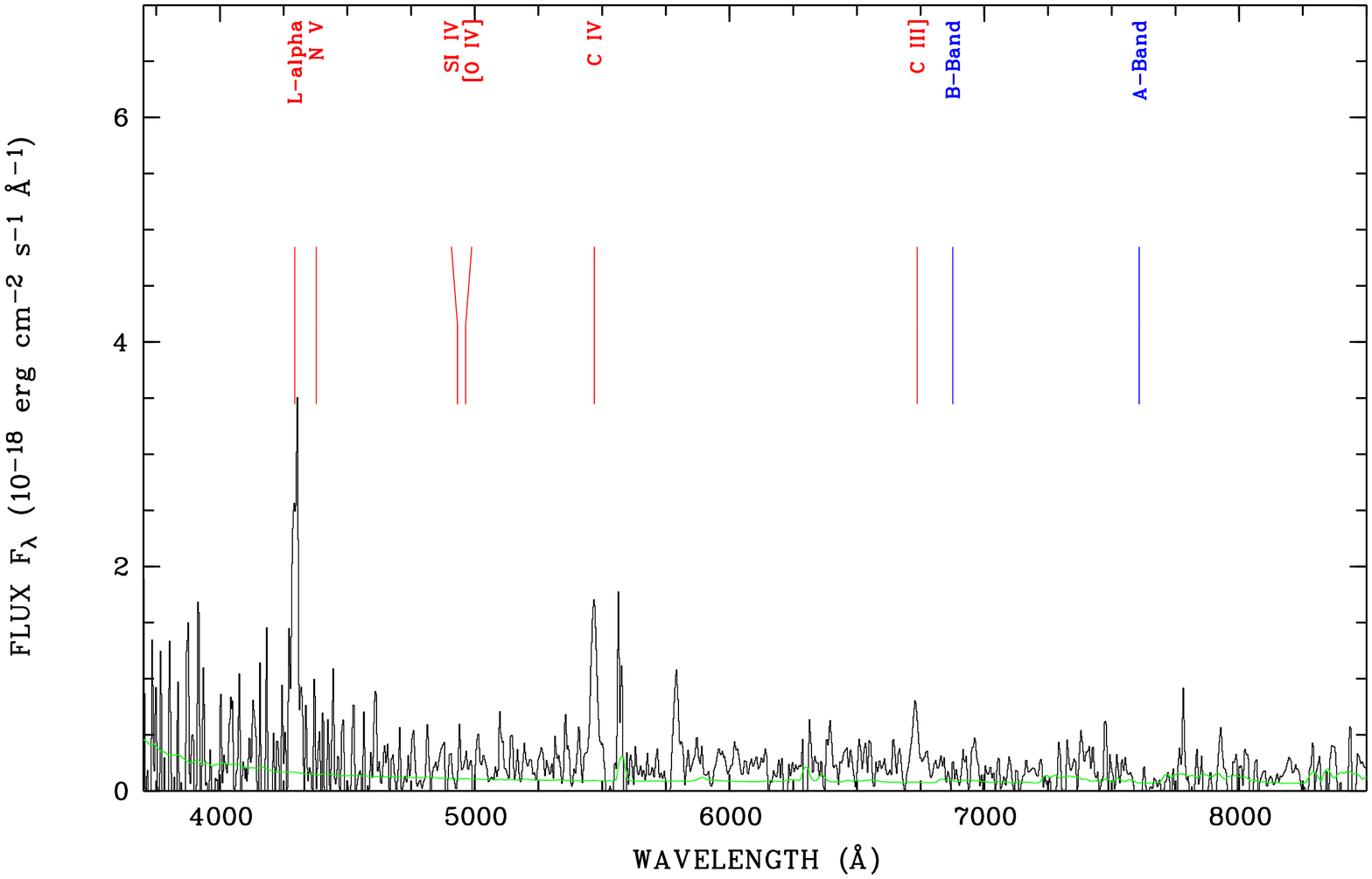,width=6.5cm,angle=270}}
 \put(0.2,3.8){\epsfig{file=spectra/Xray_spectrum_abs_463.ps,width=6cm,angle=270}}
 \put(9.2,3.8){\epsfig{file=spectra/Xray_contour_463.ps,width=6cm,angle=270}}
 \put(4.9,-2.6){object Marano 463A,  z = 2.531, frozen $\Gamma=2.0$ X-ray spectrum fit}
\end{picture} 

\newpage
\begin{picture}(10,23)
 \put(4.2,23){\epsfig{file=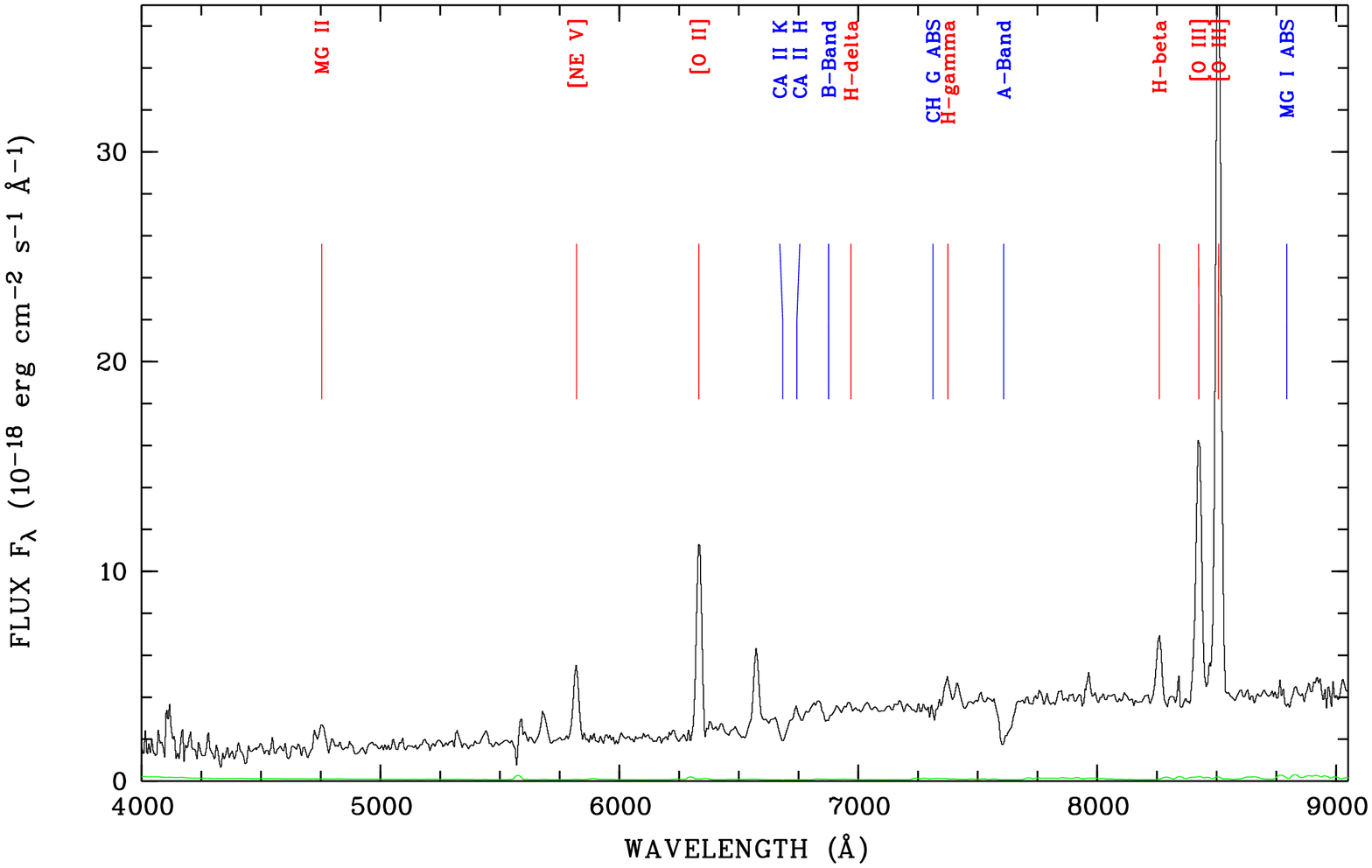,width=6.5cm,angle=270}}
 \put(0.2,16.8){\epsfig{file=spectra/Xray_spectrum_pow_zwabspow_610.ps,width=6cm,angle=270}}
 \put(9.2,18.5){\epsfig{file=spectra/Xray_contour_pow_zwabspow_610.ps,width=7.5cm,angle=270}}
 \put(3.9,10.4){object Marano 610A,  z = 0.699, two power laws, frozen $\Gamma=2.0$ X-ray spectrum fit}

 \put(4.2,10){\epsfig{file=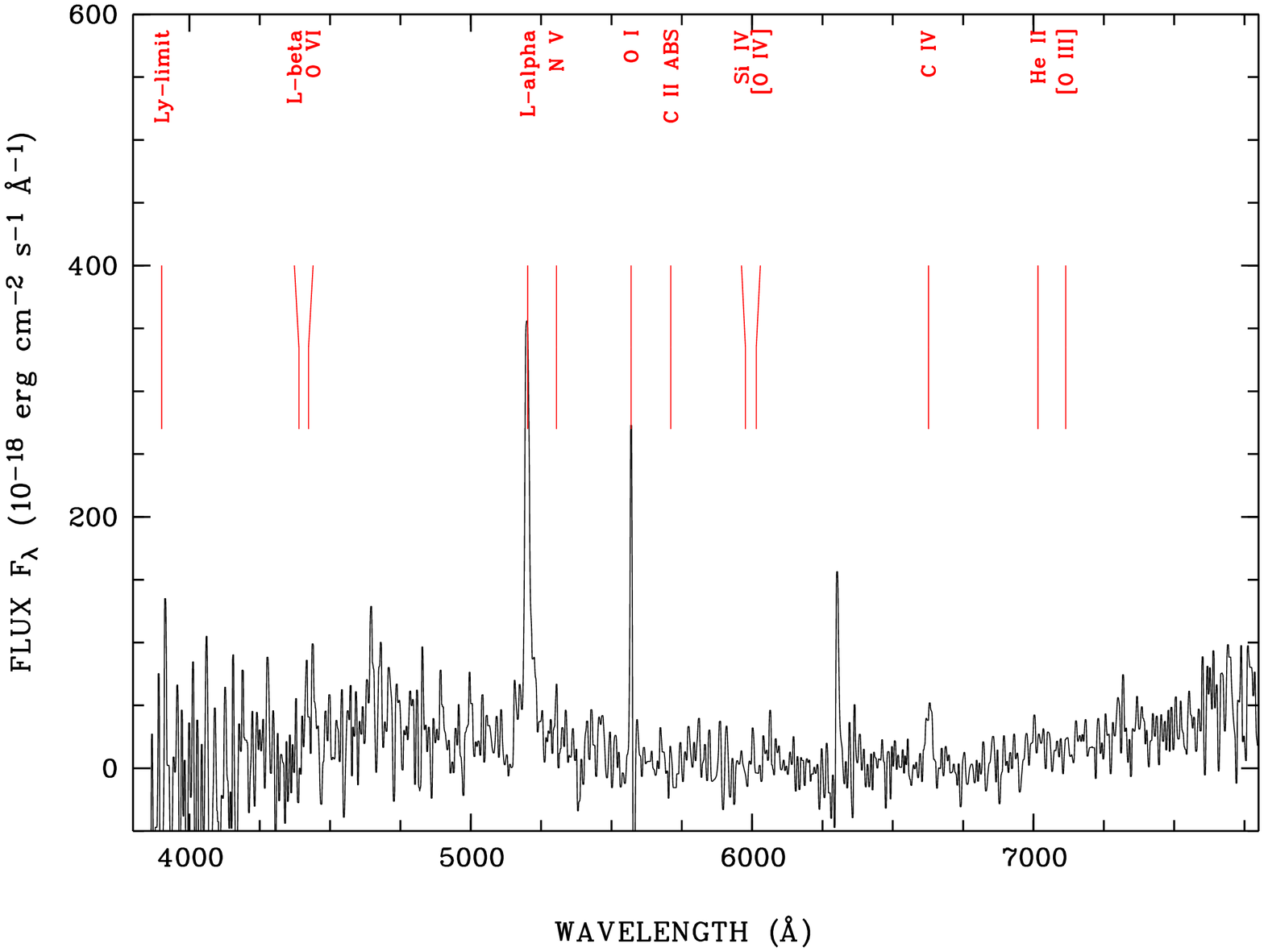,width=6.5cm,angle=270}}
 \put(0.2,3.8){\epsfig{file=spectra/Xray_spectrum_abs_X21516.ps,width=6cm,angle=270}}
 \put(9.2,3.8){\epsfig{file=spectra/Xray_contour_X21516.ps,width=6cm,angle=270}}
 \put(4.9,-2.6){object X21516\_135,  z = 3.278, frozen $\Gamma=2.0$ X-ray spectrum fit}
\end{picture} 

\newpage
\begin{picture}(10,23)
 \put(4.2,23){\epsfig{file=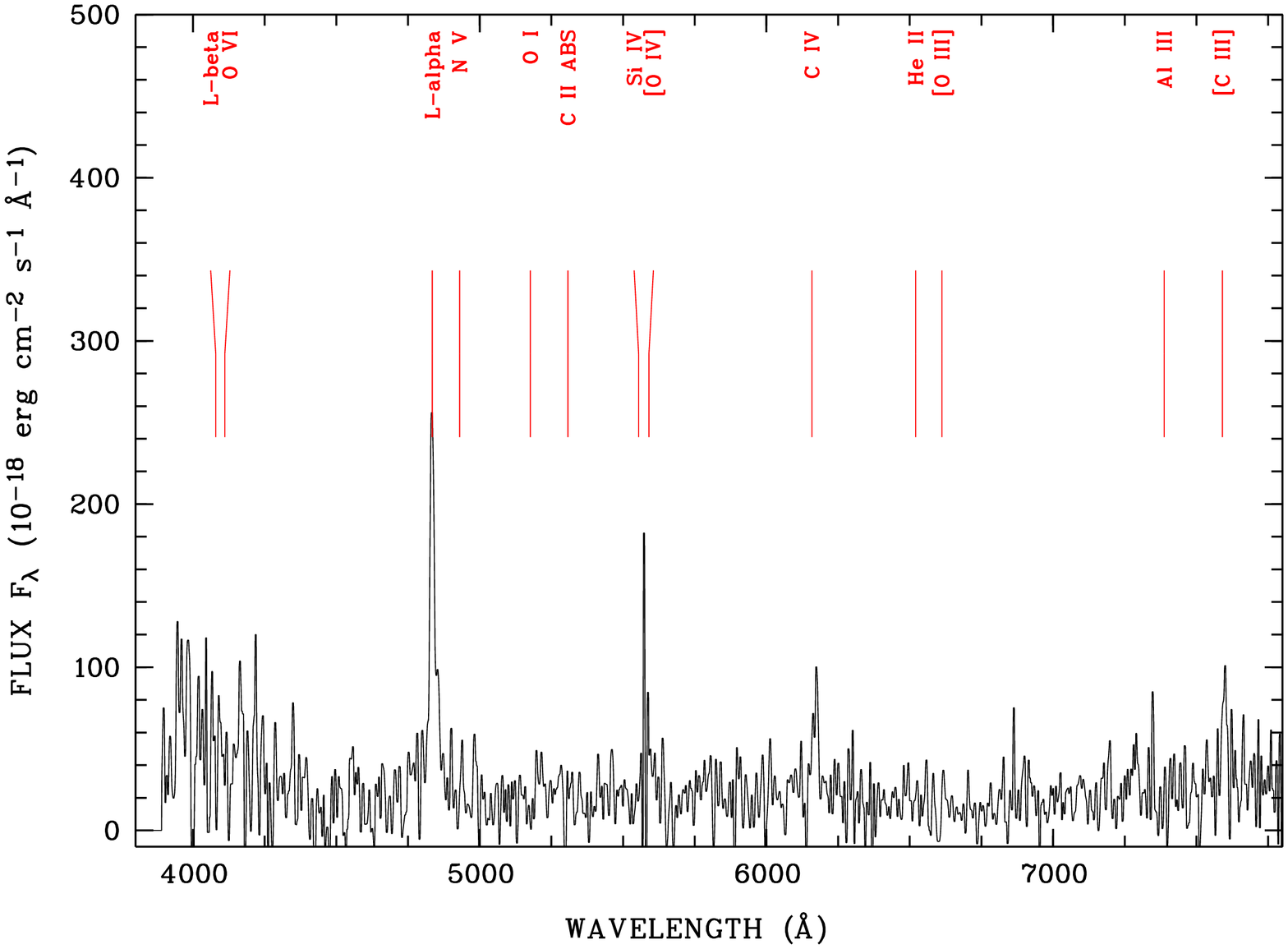,width=6.5cm,angle=270}}
 \put(0.2,16.8){\epsfig{file=spectra/Xray_spectrum_abs_X00851.ps,width=6cm,angle=270}}
 \put(9.2,16.8){\epsfig{file=spectra/Xray_contour_X00851.ps,width=6cm,angle=270}}
 \put(5.4,10.4){object X00851\_154,  z = 2.976, free X-ray spectrum fit}

 \put(4.2,10){\epsfig{file=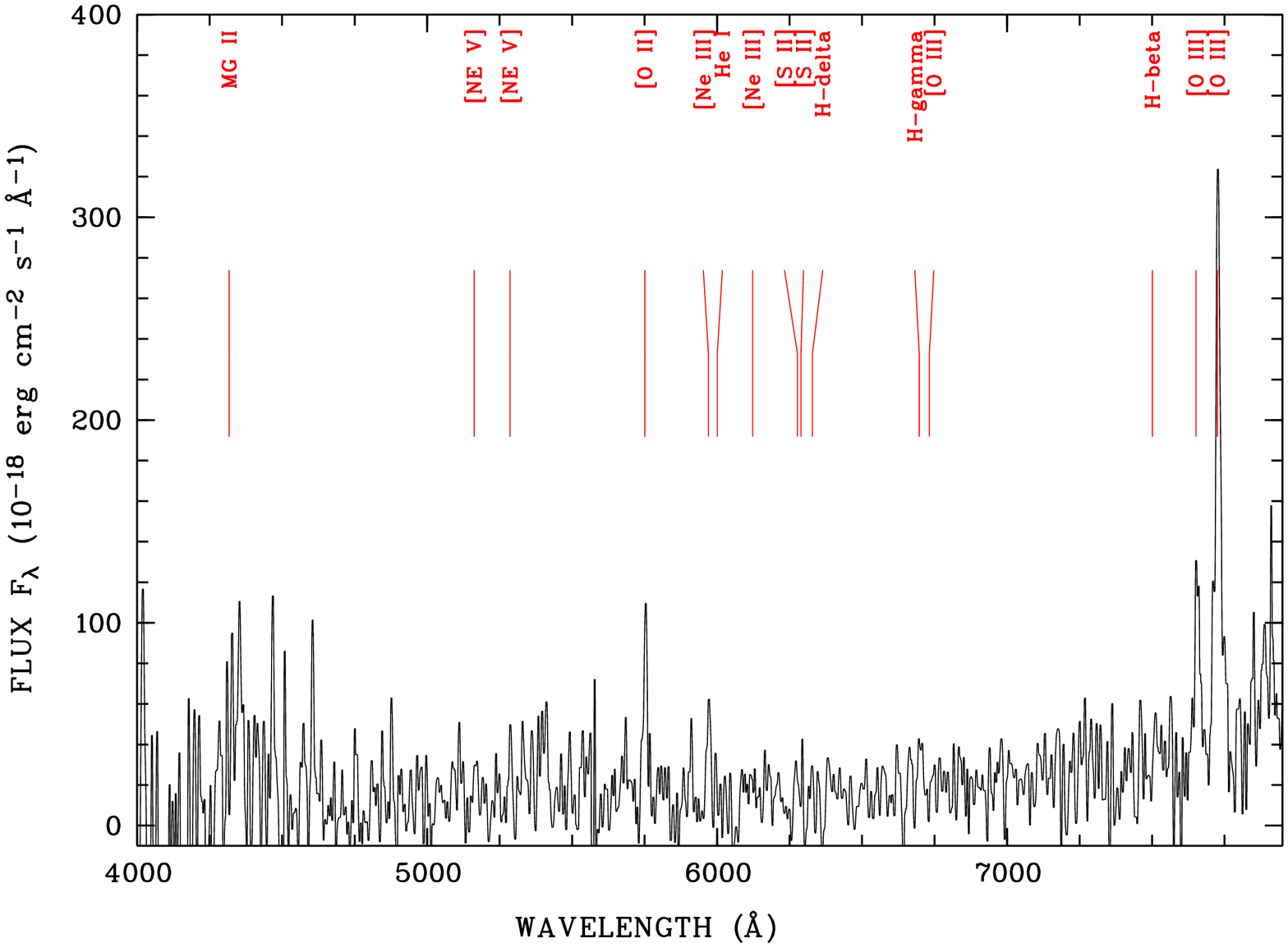,width=6.5cm,angle=270}}
 \put(0.2,3.8){\epsfig{file=spectra/Xray_spectrum_abs_X01135.ps,width=6cm,angle=270}}
 \put(9.2,3.8){\epsfig{file=spectra/Xray_contour_X01135.ps,width=6cm,angle=270}}
 \put(5.4,-2.6){object X01135\_126,  z = 0.543, free X-ray spectrum fit}
\end{picture} 

\newpage
\begin{picture}(10,23)
 \put(4.2,23){\epsfig{file=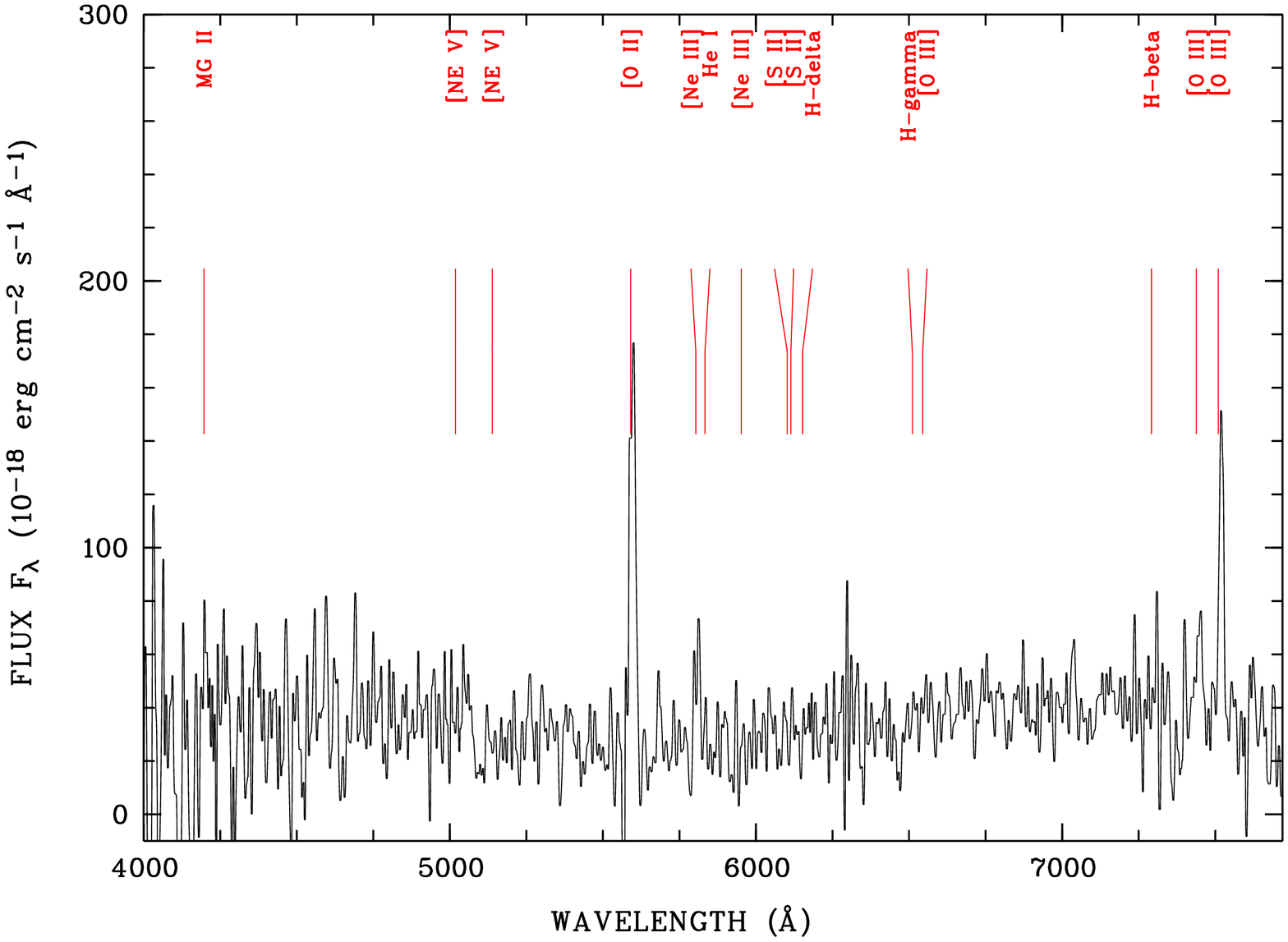,width=6.5cm,angle=270}}
 \put(0.2,16.8){\epsfig{file=spectra/Xray_spectrum_abs_X03246.ps,width=6cm,angle=270}}
 \put(9.2,16.8){\epsfig{file=spectra/Xray_contour_X03246.ps,width=6.9cm,angle=270}}
 \put(4.1,10.4){object X03246\_092, z = 0.500, frozen $\Gamma=2.0$ X-ray
 spectrum fit (0.3-8\,keV)}

 \put(4.2,10){\epsfig{file=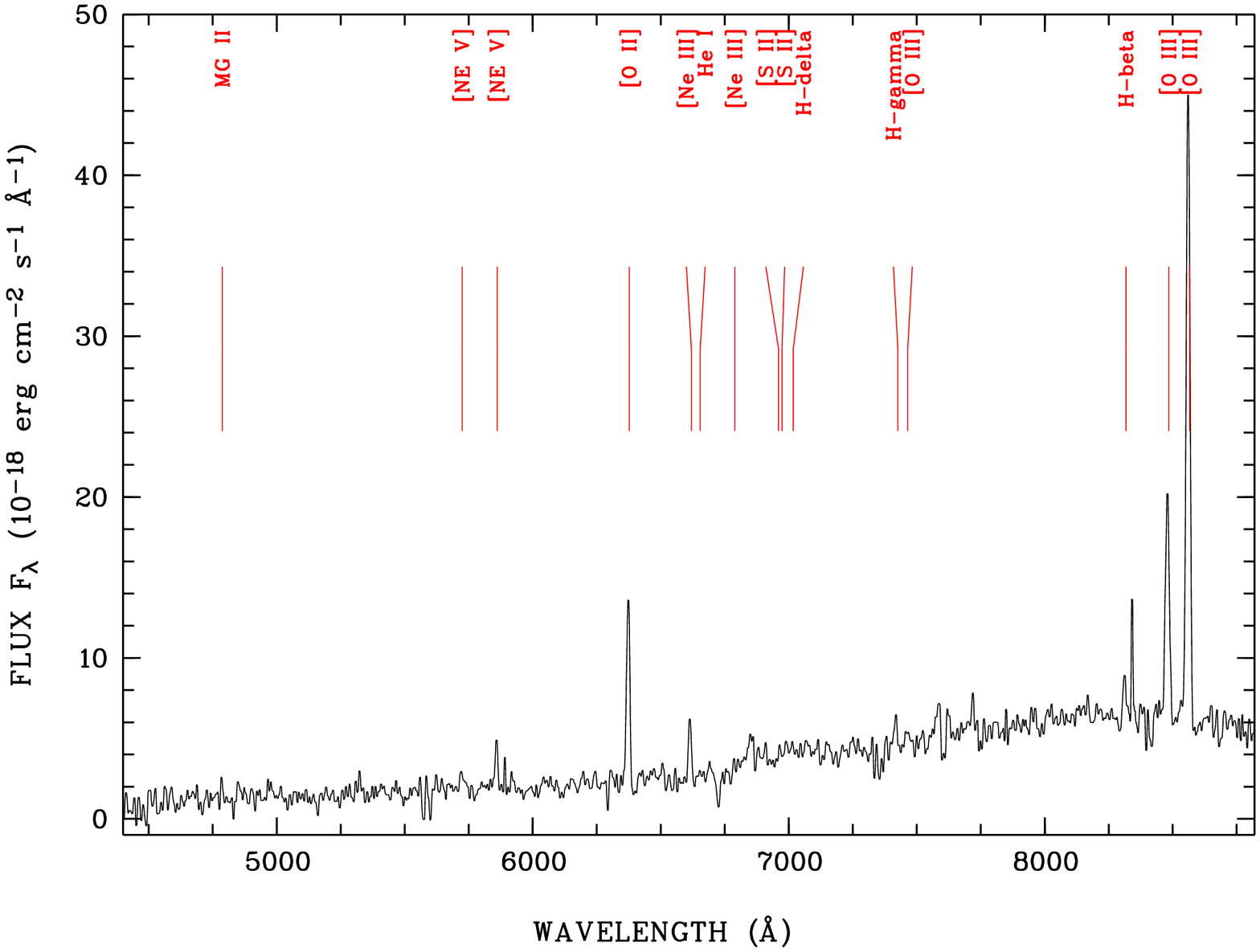,width=6.5cm,angle=270}}
 \put(0.2,3.8){\epsfig{file=spectra/Xray_spectrum_abs_phl5200.ps,width=6cm,angle=270}}
 \put(9.2,3.8){\epsfig{file=spectra/Xray_contour_phl5200.ps,width=6cm,angle=270}}
 \put(5.4,-2.6){object phl5200-001,  z = 0.711, free X-ray spectrum fit}
\end{picture}

\newpage
\begin{picture}(10,23)
 \put(3.0,17.0){\epsfig{file=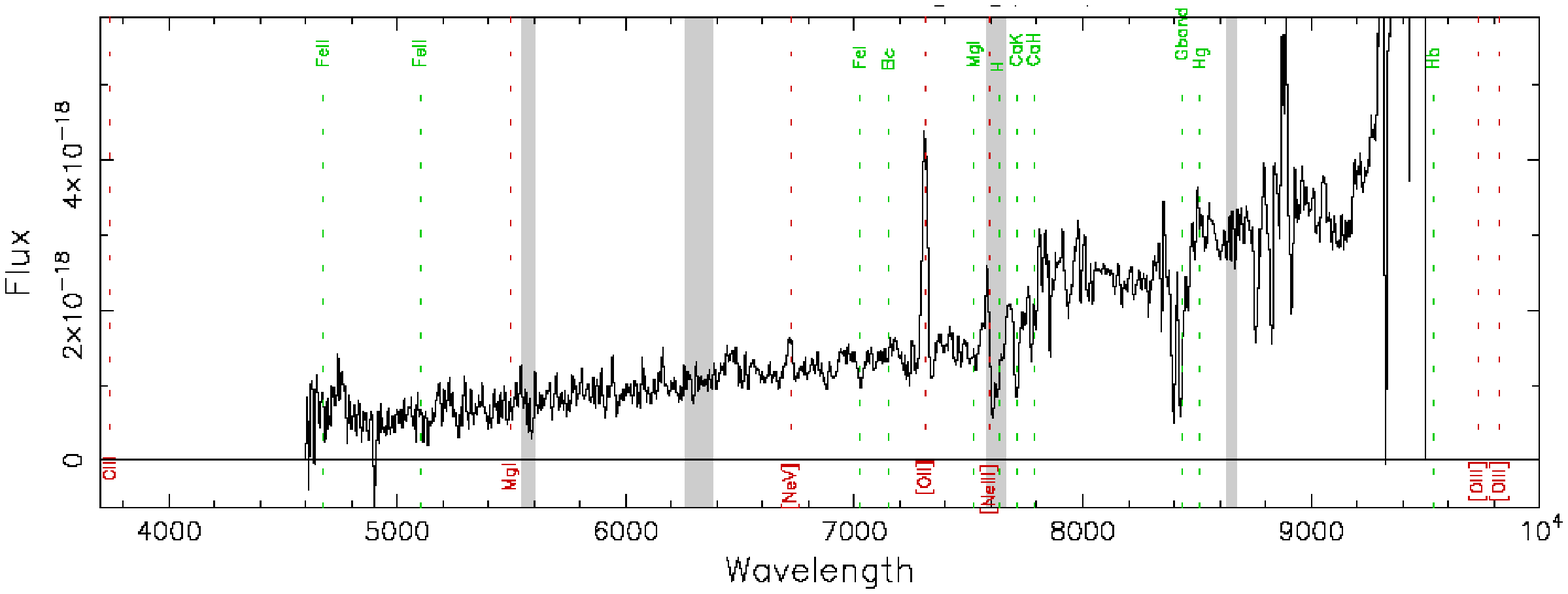,width=10cm,angle=0}}
 \put(0.2,16.6){\epsfig{file=spectra/Xray_spectrum_abs_sds1b.ps,width=6cm,angle=270}}
 \put(9.2,16.6){\epsfig{file=spectra/Xray_contour_sds1b.ps,width=6cm,angle=270}}
 \put(5.4,10.2){object sds1b-014, z = 0.962, free X-ray spectrum fit}
 \put(5.5,9.7){(optical spectrum taken from Akiyama et al. 2008)}
\end{picture} 

\end{document}